\documentclass[12pt]{article}

\usepackage[dvips]{graphicx}
\usepackage{amsmath}
\usepackage{amsfonts}

\usepackage[a4paper]{typearea} % AusgangspapiergrM-vM-_e
  \areaset[0cm]%                 % ZusM-dtzlicher Rand fuer die Bindung
          {17.0cm}{24cm}           % Textbreite und HM-vhe
  \newcommand{\PBfootnote}[1]{ }
  \newcommand{\PB}[1]{}
  \newcommand{\PBtxt}[1]{}

  \newcommand{\HIDDEN}[1]{}

  \newcommand{\PSImagxy}[3]{\includegraphics[width=#2,height=#3]{psplots/#1.gz}} 
  \newcommand{\PSImagx}[2]{\includegraphics[width=#2]{psplots/#1.gz}}

\newcommand{\BILD}[4]{\begin{figure}[#1]%

     \centerline{\parbox{15cm}{\caption[.]{#3} \label{#4}}}
     \end{figure} }

\newcommand{\Caption}[2]{%
     \centerline{\parbox{15cm}{\caption[.]{#1} \label{#2}}}}

\newcommand{\figref}[1]{fig.~\ref{#1}}
\newcommand{\tabref}[1]{table~\ref{#1}}
\newcommand{\Int}{\int\limits}
\newcommand{\IInt}{\iint\limits}

\newcommand{\e}{\operatorname{e}}
\newcommand{\ud}{\text{d}}
\newcommand{\ui}{\text{i}}
\newcommand{\ue}{\text{e}}
\newcommand{\Vol}{\operatorname{vol}}

\newcommand{\R}{\mathbb{R}}
\newcommand{\Z}{\mathbb{Z}}
\newcommand{\N}{\mathbb{N}}

\newcommand{\qlim}{\mu }

\newcommand{\slim}{\nu }
\newcommand{\slimbb}{\nu_{\text{bb}} }

\newcommand{\efsum}{\Psi_E}

\newcommand{\ALPHA}{a}

\newcommand{\setsep}{ \;\; | \;\;}

\newcommand{\OpA}{\text{A}}
\newcommand{\Op}[1]{\text{Op}[#1]}

\newcommand{\W}{\operatorname{W}}

\newcommand{\EE}{\Sigma_E}
\newcommand{\Ee}{\Sigma_1}

\newcommand{\cl}{\text{cl}}

\begin{document}

%%%%%%%%%%%%%%%%%%%%%%%%%%%%%%%%%%% title page %%%%%%%%%%%%%%%%%%%%%%%%%%%%%

\vspace*{-1cm}

%Version.: \versionsnr 
%\hfill (TeXed \today)

ULM--TP/97--8

September 1997

%%%%%%%%%%%%%
\newcommand{\titel}{On the rate of quantum ergodicity \\ \vspace*{1.25ex} in Euclidean billiards}
%%%%%%%%%%%%%

\normalsize

\vspace{0.5cm}

\renewcommand{\thefootnote}{\fnsymbol{footnote}}

\begin{center}  \huge  \bf

      \titel
%\footnote[1]{Supported by Deutsche Forschungsgemeinschaft under Contract
%             No. DFG-Ste 241/7-2} 
\end{center}

\setcounter{footnote}{1}

\begin{center}
   \vspace{3ex}
 
         {\large A.\ B\"acker%
\footnote{E-mail address: {\tt baec@physik.uni-ulm.de}}$^{1)}$, 
          R.\ Schubert%
\footnote{E-mail address: {\tt schub@physik.uni-ulm.de}}$^{1)}$, 
          and P.\ Stifter%
\footnote{E-mail address: {\tt stif@physik.uni-ulm.de}}$^{2)}$
         }

   \vspace{4ex}

   1) Abteilung Theoretische Physik\\
   2) Abteilung Quantenphysik\\ 
   Universit\"at Ulm\\
   Albert-Einstein-Allee 11\\
   D--89069 Ulm\\
   Federal Republic of Germany\\

\end{center}

\renewcommand{\thefootnote}{\arabic{footnote}}
\setcounter{footnote}{0}

%\vspace{1cm}

\vspace{0.25cm}
\begin{center}
\fbox{\begin{minipage}{15cm}
%{\Large \bf \underline{Note:}} \\
\vspace*{8pt}

{\centerline{The postscript file of this paper containing all figures
is available at: }}

\vspace*{3pt}
{\centerline{{\tt http://www.physik.uni-ulm.de/theo/qc/}}}

\vspace*{8pt}
\end{minipage}}
  \vspace{0.5cm}
\end{center}

%%%%%%%%%%%%%%%%%%%%%%%%%%%%%%%%%%%%%%%%%%%%%%%%%%%%%%%%%%%%%%%%%%%%%%%%%%%%%

%\input rate_text.tex

\newcommand{\ovl }{\bar}

\vspace*{1cm}

\leftline{\bf Abstract:}

{\small For a large class of quantized ergodic flows the quantum 
ergodicity theorem due to Shnirelman, Zelditch, Colin de Verdi\`ere 
and others states that almost all eigenfunctions become equidistributed
in the semiclassical limit.
In this work we first give a short introduction to the formulation 
of the quantum ergodicity theorem for general observables  in terms
of pseudodifferential operators and
show that it is  equivalent to  
the semiclassical eigenfunction hypothesis for the Wigner
function in the case of ergodic systems.
Of great importance is the rate by which the quantum mechanical expectation
values of an observable tend to their mean value. This 
is studied numerically for three Euclidean billiards (stadium, cosine 
and cardioid billiard) using up to 6000 eigenfunctions.
We find that in configuration space the rate of quantum ergodicity 
is strongly influenced by localized 
eigenfunctions like bouncing ball modes or scarred eigenfunctions.
We give a detailed discussion and explanation of
these effects using a simple but powerful model.
For the rate of quantum ergodicity in momentum space
we observe a slower decay. We also study the suitably normalized 
fluctuations of the expectation values around their mean, and find 
good agreement with a Gaussian distribution.  
}

\newpage

%%%%%%%%%%%%%%%%%%%%%%%%%%%%%%%%%%%%%%%%%%%%%%%%%%%%%%%%%%%%%%%%%%%%%%%%%%%%%%%
\section{Introduction}
%%%%%%%%%%%%%%%%%%%%%%%%%%%%%%%%%%%%%%%%%%%%%%%%%%%%%%%%%%%%%%%%%%%%%%%%%%%%%%%

In quantum chaos a lot of work is devoted to the investigation
of the statistics of eigenvalues and properties of eigenfunctions
of quantum systems whose classical counterpart is  chaotic.
For ergodic systems the behavior of almost all eigenfunctions in the 
semiclassical limit is described by the quantum ergodicity theorem, 
which was proven in 
\cite{Shn74,Shn93,Zel87,CdV85,HelMarRob87}, see also
\cite{Sar95,KnaSin97} for general introductions.
Roughly speaking, it states that for almost all 
eigenfunctions the expectation values of a certain class of 
quantum observables tend  to the  mean value of the corresponding 
classical observable in the semiclassical limit.

Another commonly used description of a quantum mechanical state 
is the Wigner function \cite{Wig32},
which is a phase space representation of the wave function. 
According to the  ``semiclassical eigenfunction hypothesis'' 
the Wigner function concentrates in the semiclassical limit
on regions in phase space, which
a generic orbit explores in the long time limit $t\to\infty$ 
\cite{Vor76,Vor77,Ber77b,Ber83}.
For integrable systems the Wigner function $W(p,q)$
is expected to localize
on the invariant tori, whereas for ergodic systems the Wigner
function should semiclassically condense on the energy surface, i.e.\ 
$W(p,q) \sim \frac{1}{\Vol (\EE )}\,\delta(H(p,q)-E)$,
where $H(p,q)$ is the Hamilton function and $\Vol (\EE )$ is the volume 
of the energy shell defined by $H(p,q)=E$.

As we will show below the quantum ergodicity theorem  
is equivalent to the 
 validity of the semiclassical eigenfunction hypothesis
for almost all eigenfunctions if the classical system 
is ergodic. Thus a weak form of the  semiclassical eigenfunction
hypothesis is proven for ergodic systems.

For practical purposes it is important to know not only the 
semiclassical limit of expectation values or Wigner functions, but also 
how fast this limit is achieved, because in applications one usually
has to deal with finite values of $\hbar$, or finite energies, respectively. 
Thus the  so-called rate of quantum ergodicity determines the practical 
applicability of the quantum ergodicity theorem. 
A number of articles have been devoted to this subject, see
e.g.~\cite{Zel94a,Zel94b,LuoSar95,
Sar95,EckFisKeaAgaMaiMue95} and references therein. 
The principal aim of this paper is to 
investigate the rate of quantum ergodicity numerically for different 
Euclidean billiards, and to compare the results with the 
existing analytical results and conjectures.                 
A detailed numerical analysis of the rate of quantum ergodicity
for hyperbolic surfaces and billiards can be found in \cite{AurTag97:p}.

Two problems arise when one wants to study the rate of quantum ergodicity
numerically. 
First the fluctuations of the expectation values around their mean 
can be so large that it is hard or even impossible to infer 
a decay rate. This problem can be overcome by studying 
the cumulative fluctuations, 
\begin{equation}
   S_1(E,A) = \frac{1}{N(E)} \sum_{E_n\le E} \left |\langle
   \psi_n , A\psi_n \rangle - \overline{\sigma (A)} \right | \;\;,
\end{equation}
where $\langle\psi_n , A\psi_n \rangle$ is the expectation
value of the quantum observable $A$,  
$\overline{\sigma (A)}$
is the mean value of the corresponding classical observable $\sigma (A)$
and $N(E)$ is the spectral staircase function,
see section \ref{sec:quantum-ergodicity} for detailed definitions.
So $S_1(E,A)$ contains all information about the rate by which the 
quantum expectation values tend to the mean value, but is a much 
smoother quantity than the sequence of differences itself.

Secondly, since the quantum ergodicity theorem makes only a statement about 
almost all eigenfunctions (i.e.\ a subsequence of density one, see below), 
there is the possibility of not quantum-ergodic subsequences of 
eigenfunctions. 
Such eigenfunctions can be for example  so-called scarred eigenfunctions 
\cite{Hel84,McDKau88}, which are 
localized around unstable periodic orbits, or 
in billiards with two parallel walls so-called 
bouncing ball modes, which are localized on the 
family of bouncing ball orbits.

Although such subsequences of exceptional eigenfunctions are
of density zero, they may have a considerable influence 
on the behavior of $S_1(E,A)$.
This is what we find in our numerical computations 
for the cosine, stadium and cardioid billiard, which
are based  on 2000 eigenfunctions for the cosine billiard 
and up to 6000 eigenfunctions for 
the stadium and cardioid billiard.

In order to obtain a quantitative understanding of
the influence of not quantum-ergodic subsequences on
the rate, we develop a simple model for $S_1(E,A)$  which is tested
successfully for the corresponding billiards.
The application of this model in the case of the stadium billiard
reveals in addition to the bouncing ball modes a
subsequence of eigenfunctions, which appear to be not quantum-ergodic 
in the considered energy range.

A further interesting question is if the boundary conditions have any 
influence on the rate of quantum ergodicity. This is indeed the case, 
for observables located near the boundary a strong influence on the 
behavior of 
$S_1(E,A)$ is observed. But for  $E\to\infty$ this influence vanishes, so 
the asymptotic rate is independent of the boundary conditions. 

After having some knowledge on the rate
by which the expectation values $\langle \psi_n,A\psi_n\rangle$ 
tend to their quantum-ergodic limit $\overline{\sigma (A)}$ one  is 
interested how the suitably normalized fluctuations  
$\langle \psi_n,A\psi_n\rangle -\overline{\sigma (A)}$ are distributed. 
It is conjectured that they obey a  Gaussian distribution, 
which we can confirm from our numerical data.

The outline of the paper is as follows.
In section \ref{sec:quantum-ergodicity} we first give a short introduction to  
the quantum ergodicity theorem
and its implications. Then we discuss  
conjectures and theoretical arguments 
for the rate of quantum ergodicity 
given in the literature. In particular we study the influence of not
quantum-ergodic eigenfunctions. 
In section \ref{sec:numerical} we give a detailed numerical study on the rate 
of quantum ergodicity for three Euclidean billiard systems for 
different types of observables, both in position and in 
momentum space. This includes a study of the influence 
of the boundary and a study of the fluctuations of the 
normalized expectation values around their mean.
We conclude with a summary.
Some of the more technical considerations using 
pseudodifferential operators are given in the appendix.

%%%%%%%%%%%%%%%%%%%%%%%%%%%%%%%%%%%%%%%%%%%%%%%%%%%%%%%%%%%%%%%%%%%%%%%%%%%%%%%
\section{Quantum ergodicity} \label{sec:quantum-ergodicity}
%%%%%%%%%%%%%%%%%%%%%%%%%%%%%%%%%%%%%%%%%%%%%%%%%%%%%%%%%%%%%%%%%%%%%%%%%%%%%%%

The classical systems under consideration are given by the free 
motion of a point particle inside 
a compact two--dimensional Euclidean domain $\Omega\subset \R^2$ 
with piecewise smooth boundary, where the particle 
is elastically reflected. 
The phase space is 
 given by $\R^2\times \Omega $, and the Hamilton function 
is (in units $2m=1$)
\begin{equation}
H(p,q)=p^2 \,\, .
\end{equation}
The trajectories of the flow generated by $H(p,q)$ 
lie on surfaces of constant energy $E$,  
\begin{equation}
  \EE := \left\{ (p,q) \in \R^2\times \Omega \setsep 
              p^2=E \right\} \;\; ,
\end{equation}
which obey 
the scaling property 
$\EE=E^{\frac{1}{2}}\Ee :=\{ (E^{\frac{1}{2}}p,q)\setsep (p,q)\in \Ee\}$ 
since the Hamilton function is quadratic in $p$. Note that 
$\Ee$ is just $S^1\times \Omega$.

The classical observables are functions on phase space $\R^2\times \Omega$, 
and the mean value of an observable  $a(p,q)$ at energy $E$ is given by
\begin{equation}\label{mval}
  \overline{a}^E=\frac{1}{\Vol(\EE)} 
       \Int_{\EE} a(p,q) \;\ud\mu 
        = \frac{1}{\Vol(\EE)} 
              \IInt_{\R^2\times \Omega}a(p,q) \, \delta(p^2-E) \;\ud p\,\ud q
      \;\;,
\end{equation}
where $\ud\mu=\tfrac{1}{2} \,\ud\varphi\,  \ud q $ 
is the Liouville measure on $\EE$
and $\Vol(\EE)=\int_{\EE}  \;\ud\mu$. 
The unusual factor $1/2$ in the 
Liouville measure is due to the fact that we have chosen $p^2$ and not 
$p^2/2$ as Hamilton function. 
For the mean value at energy $E=1$ we will  for simplicity write 
$\overline{a}$.

The corresponding quantum system  which we will study 
is given by the Schr\"odinger
equation
(in units $\hbar=2m=1$)
\begin{equation}
  - \Delta \psi_n(q) = E_n \psi_n(q) \;\;, \quad q\in \Omega \;\;,
\end{equation}
with Dirichlet boundary conditions: $\psi_n(q)=0$ for $q\in\partial\Omega$.
Here 
$\Delta=\frac{\partial^2}{\partial q_1^2}+\frac{\partial^2}{\partial q_2^2}$ 
denotes the usual Laplacian, and 
we will assume that the eigenvalues are ordered as 
$E_1\leq E_2\leq E_3\ldots$
and that the eigenfunctions are normalized, 
$\int_\Omega |\psi_n(q)|^2 \;\ud q = 1$.

The quantum ergodicity theorem describes the behavior of expectation
values $\langle \psi_n , A  \psi_n \rangle$ in
the high energy (semiclassical) limit $E_n\to\infty$, and relates it to the 
classical mean value (\ref{mval}).
The observable $A$ is assumed to be a pseudodifferential
operator, so before we state the theorem we have to 
introduce the concept of pseudodifferential operators,
see e.g.\ \cite{Hoe85a,Fol89,Tay81,Sch96:Diploma}. 

%%%%%%%%%%%%%%%%%%%%%%%%%%%%%%%%%%%%%%%%%%%%%%%%%%%%%%%%%%%%%%%
\subsection{Weyl quantization and pseudodifferential operators}
%%%%%%%%%%%%%%%%%%%%%%%%%%%%%%%%%%%%%%%%%%%%%%%%%%%%%%%%%%%%%%%

It is well known that every continuous operator 
$A:C^{\infty}_0(\Omega )\to \mathcal{D}'(\Omega )$ 
is characterized by its Schwarz kernel 
$K_A\in \mathcal{D}' (\Omega \times \Omega )$ 
such that $A\psi (q)=\int_{\Omega}K_A(q,q')\psi (q') \, \ud q' $,
where $\mathcal{D}'(\Omega )$ is the space of distributions dual to 
$C^{\infty}_0(\Omega )$, see e.g.\ \cite[chapter 5.2]{Hoe83a}.
In Dirac notation one has $K_A(q,q')=\langle q|A|q'\rangle$. 
With such an operator $A$ one can associate its Weyl symbol, 
defined as 
\begin{equation}\label{Weylfunction}
\W [A] (p,q):=\Int_{\R^2}\ue^{\ui q'p}\, 
K_A\left( q-\frac{q'}{2}, q+\frac{q'}{2}\right)
\, \ud q' \;\;,
\end{equation}
which in general is a distribution \cite{Fol89}. 
An operator $A$ is called a 
pseudodifferential operator, if its Weyl symbol belongs 
to a certain class of functions. One of the simplest classes of symbols is 
$S^m(\R^2\times \Omega)$ which is defined as follows: 
$a(p,q)\in S^m(\R^2\times \Omega)$, if 
it is in $C^{\infty}(\R^2\times \Omega )$ and 
for all multiindices $\alpha$, $\beta$ the estimate 
\begin{equation}\label{defsm}
\left|\frac{\partial^{|\alpha |}}{\partial p^{\alpha}}
\frac{\partial^{|\beta |}}{\partial q^{\beta}}a(p,q) \right|\leq 
C_{\alpha ,\beta}\left(1+|p|^2\right)^{\frac{m-|\alpha |}{2}} 
\end{equation}
holds. Here $m$ is called the order of the symbol. The main point in this 
definition is that  differentiation with respect to $p$ 
lowers the order of the symbol. 
For instance polynomials 
of degree $m$ in $p$, $\sum_{|\alpha ' |\leq m}c_{\alpha '}(q)p^{\alpha '}$, 
whose coefficients satisfy 
$\left|\frac{\partial^{|\beta |}}{\partial q^{\beta}} c_{\alpha '}(q)\right| \leq C_{\alpha ' ,\beta}$ are in $S^m(\R^2\times \Omega)$. 

An operator $A$ is called a pseudodifferential operator of order $m$, 
 $A\in S^{m}(\Omega)$, if its Weyl symbol 
belongs to the symbol class $S^m(\R^2\times \Omega)$, 
\begin{equation}
A\in S^{m}(\Omega) \,\, :\,\Longleftrightarrow \quad
\W [A] (p,q)\in S^m(\R^2\times \Omega)\,\,  .
\end{equation}
For example if the Weyl symbol is a polynomial in $p$, then 
the operator is in fact a differential operator and so  
pseudodifferential operators are generalizations of 
differential operators. 
Further examples include complex 
powers of the Laplacian, $(-\Delta)^{z/2}\in S^{\Re z}(\Omega )$, 
see \cite{See67,See69,Tay81}.

On the other hand, to any function $a\in S^m(\R^n\times \Omega)$ 
one can associate  an operator $\Op a \in S^m(\Omega)$, 
\begin{equation}
\Op a f (q):= \frac{1}{(2\pi)^2}\IInt_{\Omega\times\R^n} 
\e^{\ui (q-q')p} a\left(p, \frac{q+q'}{2}\right) 
f(q')\, \ud q' \ud p \,\, , 
\end{equation}
such that its Weyl symbol is $a$, i.e.\ $\W [\Op a]=a$. 
This association of the symbol $a$ to the operator $\OpA$ is 
called 
Weyl quantization of $a$. 

In practice one often encounters  symbols with a special structure, 
namely those which have 
an asymptotic expansion in homogeneous functions in $p$, 
\begin{equation}
a(p,q)\sim \sum_{k=0}^{\infty} a_{m-k}(p,q), \quad \text{with} 
\quad a_{m-k}(\lambda p,q)=\lambda^{m-k} a_{m-k}(p,q) \quad\text{for}\,\, 
 \lambda >0 \;\;. 
\end{equation}
Note that it is not required that $m$ is an integer, all 
$m\in\R$ are allowed. Since the degree of 
homogeneity tends to $-\infty$ this can be seen as  an expansion 
for $|p|\to\infty$; see \cite{Hoe85a,Fol89} for the 
exact definition of this asymptotic series.
These symbols
are often called classical or polyhomogeneous, and we will consider only  
operators with Weyl symbols of this type. The space of this operators 
will be denoted by $S^m_{\cl}(\Omega)$. 
If $A\in S^m_{\cl}(\Omega)$ and 
$\W (A)\sim \sum_{k=0}^{\infty}a_{m-k}$, then the leading term 
$a_m(p,q)$ is called the principal symbol of $A$ and is 
denoted by $\sigma (A)(p,q):= a_m(p,q)$. It plays a distinguished role in the 
theory of pseudodifferential operators. One reason for this is that operations 
like multiplication or taking the commutator are 
rather complicated in terms of the symbol, but simple for the principal 
symbol. For instance one has \cite{Hoe85a,Fol89}
\begin{equation}\label{prod}
\sigma (AB)=\sigma (A)\sigma (B)\;\;, \qquad 
\sigma ([A,B])=\ui\{\sigma (A) ,\sigma (B)\}\;\;, 
\end{equation}
where $\{\cdot ,\cdot \}$ is the Poisson bracket. It furthermore 
turns out that the principal symbol is a function on  phase space, 
i.e.\ has the right transformation properties under coordinate 
transformations, whereas the full Weyl symbol does not have this property. 

So every operator $A$  with 
principal symbol $\sigma(A)$ can be seen as a quantization of the classical 
observable $\sigma (A)$. The existence of different operators 
with the same principal symbol just reflects the fact that the
quantization process is not unique. Furthermore, one can show that the 
leading asymptotic behavior 
of expectation values of such operators for high energies only depends  
on the principal symbol, as it should be according to the correspondence 
principle. This is a special case of the Szeg\"o limit theorem, see 
\cite[chapter 29.1]{Hoe85b}.

One advantage of the Weyl quantization over other 
quantization procedures is that the Wigner function 
of a state $|\psi\rangle $ appears naturally as the Weyl symbol of 
the corresponding  projection operator $|\psi\rangle\langle \psi |$ 
\begin{equation}\label{def-Wigner}
\W\left[ |\psi\rangle\langle \psi |\right](p,q)=
\Int_{\R^2} \ue^{ \ui q' p}\, 
\, \overline{\psi} \left( q-\frac{q'}{2} \right) 
\psi\left(q+\frac{q'}{2}\right)\,
\ud q' \;\;.
\end{equation}
In the following we will use for a Wigner function of an eigenstate 
$\psi_n$ the simpler notation $\W_n (p,q):=
\W\left[ |\psi_n\rangle\langle \psi_n |\right](p,q)$. 
For the expectation value $\langle \psi ,A\psi \rangle $ one has the 
well known expression in terms of the Weyl symbol $\W [A]$ and 
the Wigner function $\W \left[ |\psi\rangle\langle \psi |\right]$, 
\begin{equation}\label{WignerWeylexpvalue}
\langle \psi , A\psi \rangle = \frac{1}{(2\pi)^2}
\IInt_{\Omega\times\R^2} \W [A](p,q)
\W \left[ |\psi\rangle\langle \psi |\right](p,q)\, \ud p\; \ud q \;\;. 
\end{equation}
Pseudodifferential operators of order zero have a bounded Wigner 
function, and therefore a bounded principal symbol 
$\sigma (A)$; this boundedness of the classical observable  
carries over to the operator level: the operators in $S^0(\Omega )$ 
are bounded in the $L^2$--norm.

The definition of pseudodifferential operators can be 
generalized to manifolds of arbitrary dimension, 
the preceeding formulas are then valid in 
local coordinates. The symbols of these operators  only 
live in local charts, but the principal symbols can be glued together 
to a function on the cotangent bundle $T^*\Omega$ which is the 
classical phase space% 
\footnote{If one wants to realize the semiclassical limit not as 
the high energy limit, 
but as the limit of $\hbar\to 0$
one has to incorporate $\hbar$ explicitly in the quantization procedure. 
In the framework of pseudodifferential operators this has been done by 
Voros in \cite{Vor76,Vor77}, see also \cite{Robe87,KnaSin97}.}.
 
%%%%%%%%%%%%%%%%%%%%%%%%%%%%%%%%%%%%%%%%%%%
\subsection{Quantum limits and the quantum ergodicity theorem}
%%%%%%%%%%%%%%%%%%%%%%%%%%%%%%%%%%%%%%%%%%%

In quantum mechanics the states are elements of a Hilbert space, 
or more generally linear functionals on the algebra of observables. 
In classical mechanics the pure states are points in phase space, and 
the observables are functions on phase space.  
More generally the states are measures on phase space, 
which are linear functionals on the algebra of observables. The pure states 
are then represented as delta functions. 
The eigenstates of a Hamilton operator are those which are invariant 
under the time evolution defined by $H$. In the 
semiclassical limit they should somehow converge to 
 measures on phase space which are invariant 
under the classical Hamiltonian flow. The measures which 
can be obtained as semiclassical limits of quantum eigenstates 
are called quantum limits. 

More concretely the quantum limits can be described as 
limits of sequences of Wigner functions. 
Let $\{\psi_n\}_{n\in\N}$ be an orthonormal basis of eigenfunctions of 
the Dirichlet Laplacian 
$-\Delta$, and $\{ \W_n \}_{n\in\N}$ the corresponding set of 
Wigner functions, see equation (\ref{def-Wigner}). 
We first consider expectation values for operators of order 
zero, and then extend the results to operators of arbitrary order.

Because pseudodifferential operators of order zero are bounded, 
the sequence of expectation values 
$\{\langle \psi_{n}, A\psi_{n}\rangle\}_{n\in\N}$ 
is bounded too. 
Every function $a\in C^{\infty}(\Ee )$ can be extended to a function in 
$C^{\infty}( \R^2\backslash \{ 0\}\times \Omega)$, by requiring it to be 
homogeneous of degree zero in $p$.
Via the quantization $\Op a$\footnote{Strictly speaking is $a$ not an allowed 
                                      Symbol because it is not smooth at $p=0$.                  Let $\chi (p)\in C^{\infty}(\R^2)$ satisfy 
                  $\chi (p)=0$ for $|p|\leq 1/4$ and $\chi (p)=1$ for 
                  $|p|\geq 1/2$. 
                  By multiplying $a$ with this excision function $\chi (p)$ 
                  we get a symbol $\chi a \in S^0(\R^2\times \Omega )$, 
                  whose Weyl quantization $\Op{\chi a}$ is in $S^0( \Omega )$. 
                  But the semiclassical properties of $\Op{\chi a}$ are 
                  independent of the special choice of $\chi (p)$, which can 
                  be see e.g.\ in \eqref{eq:distribution}, since $W_n$ is 
                  concentrated on the energy shell $\Sigma_{E_n}$ for 
                  $n\to\infty$. Therefore we will 
                  proceed for simplicity with $a$ instead of $\chi a$.  }
 of $a$ and equation 
(\ref{WignerWeylexpvalue}) one can view the Wigner function 
$\W_n (p,q)$ as a distribution on $ C^{\infty}(\Ee )$, 
\begin{equation}\label{eq:distribution}
 a\mapsto \langle \psi_n,\Op { a}\psi_n\rangle 
=\frac{1}{(2\pi)^2}\IInt_{\Omega\times \R^2}a(p,q)\W_n (p,q)\; 
\ud p\, \ud q \,\, .
\end{equation}
The sequence of these distributions is bounded because the 
operators $\Op{a}$ are bounded.
The accumulation points 
of $\{ \W_n (p,q)\}_{n\in\N}$ are called quantum limits  $\qlim_k(p,q)$,
and we label them by $k\in I $, where $I$ is some index-set.  
Corresponding to the accumulation points  $\qlim_k(p,q)$,  
the sequence
$\{ \W_n (p,q)\}_{n\in\N}$ can be split
into disjoint convergent subsequences 
$\bigcup_{k\in I}\{ \W_{n_j^k} (p,q)\}_{j\in\N}=\{ \W_n (p,q)\}_{n\in\N}$. 
That is, for every $k$ 
we have
\begin{equation}
\lim_{j\to\infty}
\IInt_{\Omega\times\R^2} a(p,q)\W_{n_j^k} (p,q)\, \ud p\; \ud q =
\IInt_{\Omega\times\R^2} a(p,q)\qlim_k(p,q)\, \ud p\; \ud q \,\, ,
\end{equation}
for all $a\in C^{\infty}(\Ee )$ viewed as homogeneous functions of degree zero 
on phase space.
This splitting is unique  up to a finite number of terms, in the 
sense that for two different splittings 
the subsequences belonging to the same accumulation point 
differ only by a finite number of terms.  

As has been shown in \cite{Zel90}, the quantum limits 
$\qlim_k$ are measures on 
$C^{\infty}(\Ee )$ which are invariant under the classical flow 
generated by $H(p,q)$.

One of the main questions in the field of quantum chaos is 
which classical invariant measures on $C^{\infty}(\Ee )$ 
can actually occur as quantum limits of Wigner functions. 
E.g., if the orbital measure
along an unstable periodic orbit 
occurs as quantum limit $\qlim_k$, then the corresponding  
subsequence of eigenfunctions has to show
an enhanced probability, i.e.\ scarring, along that
orbit.

Given any quantum limit $\mu_k$ one is furthermore interested  
in the counting function $N_k(E):=\# \{ E_{n_j^k}\leq E \}$ for the 
corresponding subsequence $\{\W_{n_j^k}\}_{j \in\N}$ of 
Wigner functions.  
Since the subsequence $\{\W_{n_j^k}\}_{j \in\N}$ is unique 
up to a finite number of elements, the corresponding 
counting function $N_k(E)$ is unique up to a constant.  

One should keep in mind that we 
have defined the quantum limits and their 
counting functions here with respect to one 
chosen orthonormal basis of eigenfunctions $\{\psi_n(q)\}_{n\in\N}$. 
If one takes a different orthonormal base of eigenfunctions 
$\{\tilde{\psi}_n(q)\}_{n\in\N}$, the counting functions corresponding 
to the quantum limits, or even the quantum limits themselves, 
may change. 
So when studying the set of all quantum limits one has to 
take all bases of eigenfunctions into account. 

The lift of any quantum limit from $\Ee$ to the whole 
phase space $\R^2\times \Omega$ follows  straightforward from some well known 
methods
in pseudodifferential operator theory, as 
shown in appendix \ref{app:generalizations-of-the-qet}.
For a pseudodifferential operator of order $m$, 
$A\in S^m_{\cl}(\Omega )$, one gets  for the expectation values 
\begin{equation}
\lim_{j\to\infty}E_{n_j^k}^{-m/2}
\langle \psi_{n_j^k}, A\psi_{n_j^k}\rangle = \qlim_k (\sigma (A)|_{\Ee})
= \Int_{\Ee}\sigma (A)(p,q)\qlim_k(p,q)\, \ud\mu \,\, .
\end{equation}
In terms of the Wigner functions this expression can be written as 
(see appendix \ref{app:connection-to-sc-eigenfunction-hypothesis})
\begin{equation}
\lim_{j\to\infty}E_{n_j^k}^{\frac{n}{2}}\W_{n_j^k}(E_{n_j^k}^{\frac{1}{2}}p,q)
= \qlim_k(p,q)\, \frac{\delta (H(p,q)-1)}{\Vol (\Ee )} \,\, .
\end{equation}
Without the scaling of $p$ with $\sqrt{E}$ we have
\begin{equation}\label{eq:Wigner-qlim}
\W_{n_j^k}(p,q)\sim \qlim_k(p,q)\frac{\delta (H(p,q)-E_{n_j^k})}{\Vol (\Sigma_{E_{n_j^k}} )}\,\, , 
\end{equation}
for $E_{n_j^k}\to\infty$, and $\qlim_k(p,q)$ is extended from $\Ee$ 
to the whole phase space by requiring it to be homogeneous of degree zero 
in $p$.

For ergodic systems the only invariant measure whose support has 
nonzero Liouville measure is the Liouville measure itself. 
For these systems the quantum ergodicity theorem states 
that almost all eigenfunctions have the Liouville density as quantum limit.

{\bf Quantum ergodicity theorem }\cite{ZelZwo96}:

{\it
  Let $\Omega\subset\R^2$ be a compact 2-dimensional domain with 
  piecewise smooth boundary, and let $\{\psi_n \}$ be an orthonormal set of 
  eigenfunctions of the Dirichlet Laplacian
  $\Delta$ on $\Omega$. 
  If the classical billiard flow on the energy shell 
  $\Ee=S^1\times \R^2$                   
  is ergodic, then there is a subsequence $\{n_j\}\subset\N$ of density 
  one such that 
  \begin{equation} \label{eq:qet}
    \lim_{j\to\infty} \; \langle \psi_{n_j}, A\psi_{n_j} \rangle
      = \overline{\sigma(A)} \;\; ,
  \end{equation}    
  for every polyhomogeneous pseudodifferential operator $A \in S^0(\Omega )$ 
  of  order zero, 
  whose Schwarz kernel $K_A(q,q')=\langle q | A |q'\rangle$
  has support in the interior 
  of $\Omega\times\Omega$. Here  $\sigma (A)$ is the principal symbol of A
  and $\overline{\sigma(A)}$ is its classical expectation value, 
  see eq.~(\ref{mval}).}

A subsequence $\{n_j\} \subset \N$ has density one if
\begin{equation}
  \lim_{E\to\infty} \frac{\#\{ n_j \;|\; E_{n_j} < E \}} {N(E)} = 1 \;\;,
\end{equation}
where $ N(E) := \#\{ n \;|\; E_n < E \}$
is the spectral staircase function, counting the number of energy 
levels below a given energy $E$. So almost all expectation values 
of a quantum observable tend to the  mean value of the corresponding 
classical observable.

The special situation that there is only one quantum limit, i.e.\
the Liouville measure, is called unique quantum ergodicity.
This behavior is conjectured to be true for the eigenfunctions of the 
Laplacian on a compact manifold of 
negative curvature \cite{Sar95,LuoSar95}.

We have stated
here for simplicity the quantum ergodicity theorem 
only for two dimensional Euclidean domains, but it is true in 
far more general situations. For compact Riemannian
manifolds without boundary the quantum ergodicity theorem was given
by Shnirelman \cite{Shn74},
Zelditch \cite{Zel87} and Colin de Verdi\`ere  \cite{CdV85}. 
For a certain class of manifolds with boundary 
it was proven in \cite{GerLei93},  without the 
restriction on the support of the Schwarz kernel of the operator $A$. The 
techniques of \cite{GerLei93} can possibly be used to remove these 
restrictions 
here as well, see the remarks in \cite{ZelZwo96}.
One can allow as well more general Hamilton operators; on manifolds 
without boundary 
every elliptic selfadjoint operator in $S_{\cl}^2(\Omega)$ is allowed, 
and on manifolds 
with boundary at least every second order elliptic selfadjoint 
differential operator with 
smooth coefficients is allowed. 
This includes for instance a free particle in a smooth 
potential or in a magnetic field.  
In the semiclassical setting, where the Hamilton operator and 
the observables depend explicitly on $\hbar$, a similar theorem for the 
limit $\hbar\to 0$ has been proven in \cite{HelMarRob87},
see also \cite{KnaSin97} for an introduction.

In light of the correspondence principle
the quantum ergodicity theorem appears very natural:
Classical ergodicity means that for a particle moving along a generic 
trajectory with energy $E$, the probability of finding it in 
a certain region $U\subset\EE$ of phase space is 
proportional to the volume $\Vol (U)$ of that region, 
but does not depend on the shape or location of $U$.  
The corresponding quantum observable is the quantization 
of the characteristic function $\chi_U$ of $U$, and by 
the correspondence principle one expects that the 
expectation value of this observable in the state $\psi_n$ 
tends to the classical expectation value for $E_n\to\infty$. 
And this is the content of the quantum ergodicity theorem. 

In terms of the Wigner functions $\W_n$ the theorem gives, see 
eq.~\eqref{eq:Wigner-qlim}, 
\begin{equation}
\W_{n_j}(p,q )\sim 
\frac{\delta (H(p,q)-E_{n_j})}{\Vol (\Sigma_{E_{n_j}}) } \,\, ,
\end{equation}
for $j\to\infty$, for a subsequence 
$\{n_j\}\subset\N$ of density one. 
So almost all Wigner functions become equidistributed on 
the energy shells $\Sigma_{E_{n_j}}$. 
That is, for ergodic systems the validity of the semiclassical 
            eigenfunction hypothesis for a subsequence of density one
            is equivalent to the quantum ergodicity theorem.

%%%%%%%%%%%%%%%%%%%%%%%%%%
\subsection{Some examples \label{sec:Some-examples}}
%%%%%%%%%%%%%%%%%%%%%%%%%%

As an illustration of the quantum ergodicity theorem, 
and for later use, we now consider some special  observables 
whose symbol only depends on the 
position $q$ or on the momentum $p$. 

If the symbol  only depends on the position $q$,  i.e.\ $a(p,q)=a(q)$, 
 the operator is just the multiplication operator with the function 
$a(q)$, and   
 one has 
\begin{equation}
   \label{num1}
   \langle \psi ,\OpA \psi \rangle = \langle \psi ,a \psi \rangle =
   \Int_\Omega a (q) \, |\psi (q)|^2 \; \ud q \;\; .
\end{equation}
In the special case that one wants to measure the probability of the particle 
to be in a given domain $D\subset \Omega$, 
the symbol is the characteristic function of 
$D$, i.e.\ $a(p,q) = \chi_D(q)$.
Then $\Op{\chi_D}$ is not a pseudodifferential operator, 
but nevertheless 
the quantum ergodicity theorem remains valid for this observable \cite{CdV85}. 
Since the  principal symbol is then  $\sigma (\OpA) = \chi_D $ 
we obtain for its mean value 
\begin{equation}
  \label{eq:qet-rhs}
  \overline{\sigma (\OpA )}=\frac{1}{\text{Vol} (\Sigma_1)}
   \Int_{S^1\times\Omega}  \chi_D (q) \, \ud\mu 
   = \frac{\Vol (D) }{\Vol (\Omega) }  \;\; .
\end{equation}
Thus the quantum ergodicity theorem gives for this case 
\begin{equation}
  \label{eq:qet-position}
  \lim_{j\to\infty} \Int_D |\psi_{n_j}(q)|^2 \;\ud q
        = \frac{\Vol{(D)}}{\Vol{(\Omega )}} \qquad
\end{equation}
for a subsequence $\{n_j\}\subset\N$ of density one. 
As discussed at the end of the previous section 
this is what one should expect 
from the correspondence principle.

If instead the symbol depends only on the momentum $p$, i.e.\ $a(p,q)=a(p)$, 
one obtains from  \eqref{WignerWeylexpvalue} for the expectation value 
\begin{equation}
  \langle \psi , \OpA\psi \rangle =
   \Int_{\R^2} a(p)  \, |\widehat{\psi} (p)|^2    \;  \ud  p \;\;.
\end{equation}
In the same way as in \cite{CdV85} for a characteristic function in 
position space, it follows that the quantum ergodicity theorem 
remains valid for
 the case where $a(p)=\chi_{C(\theta,\Delta\theta)}(p)$
is the characteristic function of a
circular sector in momentum space of angle $\theta$.
In polar coordinates this is given by the set 
\begin{equation}\label{circ-sector}
  C(\theta,\Delta\theta) := \left\{ (r,\varphi) \setsep  r \in \R^+, \; 
            \varphi \in[\theta-\Delta\theta/2,\theta +\Delta\theta/2] 
             \right\} \;\;.
\end{equation}
The mean value of the principal symbol 
then reduces to 
\begin{equation}
 \overline{\sigma (\OpA)} = \frac{1}{\Vol (\Ee )}
   \Int_{S^1\times\Omega}  \chi_{C(\theta,\Delta\theta)} (p) \, \ud\mu 
 = \frac{\Delta\theta }{2\pi } \;\; ,
\end{equation}
which does not depend on $\theta$.
Thus the quantum ergodicity theorem reads in the case
of a characteristic function in momentum space
\begin{equation} \label{eqn:qet-ft-version}
    \lim_{j \to \infty} \Int_{C(\theta,\Delta\theta)} 
              | \widehat{\psi}_{n_j}(p) |^2 \; \ud p 
            = \frac{\Delta\theta}{2\pi} 
\end{equation}                                    
for a subsequence $ \{n_j\}\subset\N$ of density one. 
This means that quantum 
ergodicity implies an asymptotic equidistribution of the 
momentum directions of the particle.

It is instructive to compute the observables discussed above
for certain integrable systems.

First consider a two-dimensional torus.
The eigenfunctions, labeled by the two quantum numbers $n,m\in \Z$,
read $\psi_{n,m}(x,y)=\exp (2\pi\ui n x) \exp(2\pi\ui m y)$.
Obviously, these are ``quantum-ergodic'' in position space, since 
$|\psi_{n,m}(x,y)|^2=1$, but they are not quantum-ergodic in 
momentum space. Even in position space the situation changes if 
one takes a different orthogonal basis of eigenfunctions (note 
that the multiplicities tend to infinity), see \cite{Jak97} for 
a complete discussion of the quantum limits 
on tori.  
A similar  example 
is provided by the Dirichlet or Neumann eigenfunctions
of a rectangular billiard.

The circle billiard shows a converse behavior. 
Let the  radius be one, then the eigenfunctions
are given in polar coordinates by
\begin{equation}
  \psi_{kl}(r,\phi) = N_{kl} J_l(j_{k,l} r)  \;
                     \ue^{\ui l \phi} \;\; .
\end{equation}
Here $j_{k,l}$ is the $k$--th zero of the Bessel function $J_l(x)$, $x>0$,
and $N_{kl}$ is a normalization constant. 
These eigenfunctions  do not exhibit quantum ergodicity in position space. 
But for their Fourier transforms one can show that
\begin{align}
  \Int_{C(\theta,\Delta\theta)} \left| \widehat{\psi}_{kl}(p)\right|^2 
      \; \ud p = \frac{\Delta\theta}{2\pi} \;\;, 
\end{align}
and so we have ``quantum ergodicity'' in momentum space.

A remarkable example was discussed by Zelditch \cite{Zel92}. 
He considered the 
Laplacian  on the sphere 
$S^2$.  Since 
the multiplicity of the eigenvalue $l(l+1)$ is $2l+1$,
 which tends to infinity as $l\to\infty$, 
one can choose infinitely many  orthonormal bases of eigenfunctions. 
Zelditch showed that almost all of these bases exhibit quantum ergodicity in 
the whole phase space. Although this is clearly an exceptional case 
due to the high multiplicities, 
it shows that one has to be careful with the notion of 
quantum ergodicity. In a recent work Jakobson and Zelditch \cite{JakZel97:p}
have furthermore shown that for the sphere all invariant measures 
on phase space do occur as quantum limits. One might conjecture that 
for an integrable system all classical measures which are invariant under 
the flow and all symmetries of the Hamilton function do occur as 
quantum limits.

The general  question whether quantum ergodicity for all 
orthonormal bases of eigenfunctions in the whole phase space 
implies ergodicity 
of the classical system is still open.

%%%%%%%%%%%%%%%%%%%%%%%%%%%%%%%%%%%%%%%%%%%%%
\subsection{The rate of quantum ergodicity 
        \label{sec:rate-of-quantum-ergodicity}}
%%%%%%%%%%%%%%%%%%%%%%%%%%%%%%%%%%%%%%%%%%%%%

We now come to the central question 
of the approach to the quantum-ergodic limit.
First we note that an equivalent formulation of the quantum ergodicity 
theorem, which avoids choosing subsequences, is
given by
  \begin{equation} 
    \label{eqn:qet-sum-version}
    \lim_{E \to\infty} \frac{1}{N(E)}  \sum_{E_n \le E} 
         \left| \langle \psi_n, A\psi_n \rangle 
           -\overline{\sigma(A)}  \right|= 0 \;\;.
  \end{equation}
This equivalence follows from a standard lemma concerning the influence 
of subsequences of density zero on the average of a sequence, see e.g.\  
\cite[Theorem 1.20]{Wal82}.

In order to characterize the rate of approach to the ergodic limit the
quantities 
\begin{equation}
   S_m(E,A) = \frac{1}{N(E)} \sum_{E_n\le E} \left |\langle
   \psi_n , A\psi_n \rangle - \overline{\sigma (A)} \right |^m
\end{equation}
have been proposed and studied in \cite{Zel94a,Zel94b}.
Quantum ergodicity is equivalent to  $S_m(E,\OpA) \to 0$ for $E\to\infty$ and 
$m\geq 1$.

Let us first summarize some of the known results for the rate of 
quantum ergodicity.

Zelditch proved in \cite{Zel94a}
by relating the rate of quantum ergodicity to the rate of convergence
of classical expectation values, 
and using a central limit theorem for the classical flow,
that for compact manifolds of negative curvature
$S_m(E,\OpA)=O((\log E)^{-m/2})$. However this bound is believed to be
far from being sharp. 
Moreover in \cite{Zel94b} lower bounds for $S_m(E,\OpA)$
have been derived.
In \cite{LuoSar95,Jak94,Jak97b} it is proven for a Hecke basis 
of eigenfunctions
on the modular surface that $S_2(E,A) <C(\varepsilon) E^{-\frac{1}{2}+\varepsilon}$ 
for every $\varepsilon >0$.  It is furthermore conjectured 
\cite{Sar95,LuoSar95} 
that this 
estimate is also valid for the eigenfunctions of the 
Laplacian on a compact manifolds of negative curvature, and moreover that 
it is satisfied for each eigenstate individually:
$|\langle \psi_n, A\psi_n\rangle -\overline{\sigma (A)}|<
C(\varepsilon)E^{-\frac{1}{4}+\varepsilon}$ for every 
$\varepsilon >0$. 

\newcommand{\classvar}{\rho(A)}
\newcommand{\SIIsc}{S_2}
\newcommand{\DELTA}{\eta}

In \cite{EckFisKeaAgaMaiMue95} a study of $S_2(E,A)$ based on the Gutzwiller 
trace formula has been performed. 
For completely desymmetrized systems having only isolated and unstable
periodic orbits, 
the so-called diagonal 
approximation for a double sum over periodic orbits and 
further assumptions 
lead to 
\begin{equation}\label{eq:rate-hypsyst}
  \SIIsc(E,A)\sim g\frac{2}{\Vol ( \Omega) }\, \classvar \,E^{-\frac{1}{2}} 
            \;\; .
\end{equation}
Here $g=2$ if the system is invariant under time reversal, and otherwise 
$g=1$, and $\classvar$ is the variance of the  fluctuations of 
$ A_{\gamma}=\tfrac{1}{T_{\gamma}}\int_0^{T_{\gamma}}
\sigma (A)(\gamma (t))\, \ud t$ around their mean $\overline{\sigma (A)}$, 
computed using all periodic orbits $\gamma$ of the system. More precisely,  
it is assumed that $|A_{\gamma}-\overline{\sigma (A)}|^2\sim 
\classvar /T_{\gamma}$, where $T_{\gamma}$ denotes the primitive length 
of $\gamma$.

In the general case where not all periodic orbits are isolated and unstable 
it is argued that 
the rate of quantum ergodicity is related to the 
decay rate of the classical autocorrelation function $C(\tau)$
\cite{EckFisKeaAgaMaiMue95}.
If  $C(\tau) \sim \tau^{-\DELTA}$
then the result is
\begin{equation}\label{eq:gen-rate}
  S_2(E,A) \sim \Int_0^{T_H}C(\tau )\, \ud \tau \sim \begin{cases}  E^{-1/2}  & \text{ for } \DELTA>1 \\
                                \ln \Big(\frac{\Vol (\Omega)}{2}  
E^{1/2}\Big)  E^{-1/2} 
                                          & \text{ for } \DELTA=1 \\
                                E^{-\DELTA/2}
                                          & \text{ for } \DELTA<1 
                 \end{cases} \;\; ,
\end{equation} 
where $T_H=\frac{\Vol (\Omega)}{2}  E^{1/2}$ is the so-called 
Heisenberg time. 
  
For the stadium billiard \cite{Bun74}
and the Sinai billiard \cite{Sin70}
it is believed that the correlations decay 
as $\sim 1/\tau$, see \cite{Bun85} and \cite{DahArt96}
for
numerical results for the Sinai billiard.
Thus for both the stadium and the Sinai billiard
a logarithmic contribution to the decay  
of $S_2(E,A)$ is  expected.

Also a Gaussian random behavior of the eigenfunctions \cite{Ber77b}
implies in position space 
a rate $S_2(E,A)=O(E^{-\frac{1}{2}})$,
which follows from \cite[chapter IV]{Sre94}, 
see also \cite{EckFisKeaAgaMaiMue95,SreSti96}.

Random matrix theory (see \cite[section VII]{BroFloFreMelPanWon81})
predicts for suitable observables the same 
rate $S_2(E,A)=O(E^{-\frac{1}{2}})$, and furthermore Gaussian fluctuations 
of $(\langle \psi_n, A\psi_n\rangle -\overline{\sigma(A)})/\sqrt{S_2(E_n, A)}$ 
around zero, which we study numerically in 
section~\ref{sec:fluct-of-exp-values}.

Since for the systems under investigation 
we have not quantum-ergodic subsequences of 
eigenfunctions, we now discuss in general the influence of
such subsequences
on the behavior of $S_1(E,A)$. 
To this end we split the sequence of eigenfunctions into two subsequences.
The first, denoted by $\{\psi_{n'}\}$, 
contains all quantum-ergodic eigenfunctions,
i.e.\ the corresponding quantum limit of the associated sequence
of Wigner functions is the Liouville measure.
The counting function of this subsequence will be denoted by $N'(E)$.
The other sequence $\{\psi_{n''}\}$ 
contains all not quantum-ergodic eigenfunctions.
This subsequence may have different quantum limits $\qlim_k$
which are all different from the Liouville measure.
Their counting function will be denoted by $N''(E)$. 
Examples would be a subsequence of bouncing ball modes 
or  eigenfunctions scarred by an unstable periodic orbit. 
Similarly we split $S_1(E,A)$ into two parts corresponding to the two 
classes of eigenfunctions.
Due to the separation  $N(E)=N'(E)+N''(E)$ we obtain
\begin{equation} \label{eq:split-of-S1}
\begin{split}
S_1(E,A)=\frac{1}{N(E)}\sum_{E_n\leq E}
\left|\langle \psi_n ,A\psi_n \rangle -\overline{\sigma (A)}\right|
&= \frac{N'(E)}{N(E)}S_1'(E,A)+\frac{N''(E)}{N(E)}S_1''(E,A)\\
&=\left(1-\frac{N''(E)}{N(E)}\right)S_1'(E,A)+\frac{N''(E)}{N(E)}S_1''(E,A)\,\, .
\end{split}
\end{equation}
Here we defined 
\begin{align}
S_1'(E,A)&:=\frac{1}{N'(E)}
\sum_{E_{n'}\leq E}\left|\langle \psi_{n'} ,A\psi_{n'} \rangle 
                                    -\overline{\sigma (A)}\right|\,\, ,\\
 S_1''(E,A)&:=\frac{1}{N''(E)}\sum_{E_{n''}\leq E}
\left|\langle \psi_{n''} ,A\psi_{n''} \rangle -\overline{\sigma (A)}\right|\,\, .
\end{align}
So the behavior of $S_1(E,A)$ is given in terms of the three 
quantities $S_1'(E,A)$,  $S_1''(E,A)$ and $N''(E)$, which 
describe the behavior of the quantum-ergodic and  the not quantum-ergodic  
subsequences, respectively. 

The behavior of $S_1 ''(E,A)$ can be described in terms of the 
non ergodic quantum limits 
and their counting functions.
We split the not quantum-ergodic subsequence into convergent subsequences 
corresponding to the quantum limits $\qlim_k\neq \mu $, 
$\{\psi_{n''}\}=\bigcup_k\{\psi_{n^k_j}\}_{j\in\N}$,  with 
$N''(E)=\sum_k N_k(E)$, and 
$\langle \psi_{n^k_j} ,A\psi_{n^k_j} \rangle -\overline{\sigma (A)}\sim
\qlim_k \Bigl(\sigma (A)-\overline{\sigma (A)}\Bigr)$. Then $S_1''(E,A)$ 
is asymptotically given by 
\begin{equation}
S_1''(E,A)\sim \frac{1}{\sum_k N_k(E)}\sum_k N_k(E)
\left|\qlim_k \Bigl(\sigma (A)-\overline{\sigma (A)}\Bigr)\right|\,\, , 
\end{equation}
and the limit
\begin{equation}\label{eq:nu-pp-def}
\slim ''(A):=\lim_{E\to\infty}S_1''(E,A)
\end{equation} 
only depends on $\sigma (A)$ and defines an invariant measure on $\Ee$.

Let us assume for the quantum-ergodic part of $S_1(E,A)$ a certain 
rate of decay,
\begin{equation}\label{quant-erg-rate}
S_1'(E,A)=\slim '(A)E^{-\alpha }+o(E^{-\alpha })\,\, ,
\end{equation}
and for the counting function  of the not quantum-ergodic states 
\begin{equation} \label{eq:counting-fct-non-qerg-states}
N''(E) =c E^{\beta }+o(E^{\beta }) \,\, ,
\end{equation}
where by quantum ergodicity $\alpha >0$ and $\beta<1$. 
With Weyl's law 
$N(E)=\frac{\Vol (\Omega )}{4\pi}\, E +O(E^{\frac{1}{2}})$
we then obtain in eq.~\eqref{eq:split-of-S1} for $S_1(E,A)$ 
\begin{equation}\label{eq:reduced-rate-due-to-nerg-states}
S_1(E,A)=\slim '(A)E^{-\alpha }+\frac{4\pi c}{\Vol (\Omega )}\slim ''(A)E^{\beta -1}+o(E^{-\alpha })+
o(E^{\beta -1})\,\, .
\end{equation}
One sees that if $-\alpha >\beta -1$, the asymptotic behavior 
of $S_1(E,A)$ is governed by the quantum-ergodic sequences of 
eigenfunctions, whereas in the opposite case, $-\alpha \le \beta -1$, 
the not quantum-ergodic sequences dominate the behavior asymptotically. 
Especially if $\beta -1> - 1/4$, i.e.\ $\beta>3/4$,  
the rate of quantum ergodicity 
cannot be $O(E^{-\frac{1}{4}})$. 

\newcommand{\Sm}{S_1^{\text{model}}}

To obtain a simple model for the rate of quantum ergodicity, 
let us now assume that the conjectured optimal rate 
is valid for the subsequence of quantum-ergodic 
eigenfunctions, that is $\alpha =1/4$ can be chosen
in eq.~(\ref{quant-erg-rate}). 
To be more precise it should be $S_1'(E,A)=O(E^{-1/4+\varepsilon})$ 
for every $\varepsilon >0$, but for comparison with numerical data 
on a finite energy range we will assume that $\varepsilon =0$.  
For the not quantum-ergodic eigenfunctions the knowledge 
of their counting function $N''(E)$ is very poor; in general it 
is unknown.
Thus if we neglect the higher order terms in eqs.~(\ref{quant-erg-rate})
and (\ref{eq:counting-fct-non-qerg-states}) 
we obtain from \eqref{eq:split-of-S1}
and \eqref{eq:nu-pp-def} a simple model for the behavior of $S_1(E,A)$, 
\begin{equation}\label{mod-rate}
\Sm(E,A)=\left(1-\frac{4\pi c}{\Vol(\Omega)}E^{\beta -1}\right)\,\slim '(A)E^{-\frac{1}{4}}
+\frac{4\pi c}{\Vol (\Omega )}\slim ''(A)E^{\beta -1}\,\, .
\end{equation}
The first factor in 
braces will only be important if $\beta$ is close to 1.

We will now discuss the influence of a special type of not quantum-ergodic 
subsequences in more detail. 
In billiards with two--parallel walls, one 
has a subsequence of so-called bouncing ball modes \cite{McDKau79}, 
which are localized on the bouncing ball orbits,
see fig.~\ref{fig:stad-waves}b) for an example 
of such an eigenfunction.
In our previous work \cite{BaeSchSti97a} 
we showed that for every $\beta <1$ there exists
an ergodic billiard which possesses 
a not quantum-ergodic subsequence, given by 
 bouncing ball modes, whose 
counting function is asymptotically of order $E^{\beta}$. But for 
$\beta =1-\delta$, with some small $\delta >0$, 
equation (\ref{eq:reduced-rate-due-to-nerg-states}) shows that
$S_1(E,A)=O(E^{-\delta})$ at least for some $A$. So the best 
possible estimate on the rate of quantum ergodicity 
which is valid without further assumptions on the system other than 
ergodicity  is
\begin{equation}
S_1(E,A)=o(1) \,\,, \qquad \text{i.e. } \lim_{E\to\infty} S_1(E,A) =0    \,\, .
\end{equation}
Especially for the Sinai billiard the result for the exponent 
is $\beta=9/10$ and therefore $S_1(E,A)\sim cE^{-1/10}$, which 
contradicts the 
result \eqref{eq:gen-rate} 
from \cite{EckFisKeaAgaMaiMue95}.

If the bouncing ball modes are the only not quantum-ergodic 
eigenfunctions, or at least constitute the dominant contribution 
to them, then $N''(E)\sim N_{\text{bb}}(E)\sim cE^{\beta}$. 
  The exponent 
$\beta$ and $\slim ''(A)$ are explicitly known, and the 
constant $c$ is known from a numerical fit in \cite{BaeSchSti97a} 
for the billiards we will consider in the next section. 
Thus in this case 
the only free parameter in the model \eqref{mod-rate} is $\slim '(A)$.

The asymptotic behavior of (\ref{mod-rate}) is governed 
by the term with the larger exponent, but this can be hidden 
at low  energies  if one of the constants is much larger 
than the other.
Assume for instance that $\beta -1> 1/4$, 
i.e.\ the not quantum-ergodic eigenfunctions
 dominate the rate asymptotically. If
\begin{equation}\label{frac-const}
\frac{4\pi c\slim ''(A)}{\Vol (\Omega )\slim '(A)}\ll 1\,\, ,
\end{equation}
for an observable $A$, then up to a certain energy 
$S_1(E,A)$ will be approximately proportional to $E^{-\frac{1}{4}}$. 
In  numerical studies where only a finite energy range  is 
accessible such a behavior can hide the true rate 
of quantum ergodicity. 
This will be seen most drastically for the cosine billiard, see 
section~\ref{sec:qerg-rate-cos-billiard}.

The main ingredient of the model (\ref{mod-rate}) is the 
conjectured behavior of the rate for the 
quantum-ergodic eigenfunctions. By comparing (\ref{mod-rate}) 
with numerical data for different observables one can 
test this conjecture. 
If this conjecture is true then it means that the 
only deviations from the optimal rate 
of quantum ergodicity 
are due to subsequences of not quantum-ergodic eigenfunctions.

Clearly  similar models based on a splitting like 
\eqref{eq:split-of-S1} can be developed for other situations as well.
E.g., if the eigenfunctions split into a  quantum-ergodic subsequence 
of density one 
with rate proportional to $E^{-1/4}$ and a quantum-ergodic
subsequence of  density zero with a slower, and maybe spatial 
inhomogeneous, rate, 
one would expect a similar behavior of $S_1(E,A)$ as in the case 
considered above. So it will be hard without some a priori information on 
not quantum-ergodic eigenfunctions to distinguish between these two 
scenarios.

%%%%%%%%%%%%%%%%%%%%%%%%%%%%%%%%%%%%%%%%%%%%%%%%%%%%%%%%%%%%%%%%%%%%%%%%%%%%%%%
\section{Numerical results} \label{sec:numerical}
%%%%%%%%%%%%%%%%%%%%%%%%%%%%%%%%%%%%%%%%%%%%%%%%%%%%%%%%%%%%%%%%%%%%%%%%%%%%%%%

In order to study the rate of quantum ergodicity numerically 
we have chosen three different Euclidean billiard systems,
given by the free motion of a point particle inside a compact domain
with elastic reflections at the boundaries.
See \figref{fig:billiard-domains} for the chosen 
billiard shapes.

The first is the stadium billiard, which
is proven to be ergodic, mixing and a $K$-system \cite{Bun74,Bun79}.
The height of the desymmetrized billiard 
is chosen to be 1, and $a$ denotes the length
of the upper horizontal line.
For this system our analysis is based on computations of
the first 6000 eigenfunctions 
for odd-odd parity, i.e.\
everywhere Dirichlet boundary conditions in the desymmetrized system
with parameter $a=1.8$.
We also studied  stadium billiards with parameters $a=0.5$ and $a=4.0$
using the first 2000 eigenfunctions in each case
to investigate the dependence on $a$, see below.
The stadium billiard is one of the most intensively
studied systems in quantum chaos, for
investigations of the eigenfunctions see e.g.\ 
\cite{McDKau79,Hel84,McDKau88,Li97,SimVerSar97:p}
and references therein.

The second system is the cosine billiard,
which is constructed by replacing one side of a rectangular box by 
a cosine curve. The cosine billiard has been introduced and studied
in detail in \cite{Sti93:Diploma,Sti96:PhD}.
The ergodic properties are unknown, but numerical studies
do not reveal any stability islands. If there were any
they are so small that one expects that they do not have any influence in
the energy range under consideration.
The height of the cosine billiard is 1 and the upper horizontal
line has length  2 in our numerical computations. 
The cosine is parameterized by
$B(y)=2+\frac{1}{2}(1+\cos(\pi y))$, see \figref{fig:billiard-domains}b).
For our analysis of this system we used  the first 2000 eigenfunctions 
with Dirichlet boundary conditions everywhere.

\newcommand{\einstadbild}[2]{
  \vspace*{1.5ex}
  \PSImagx{stad_wav_#1.ps}{8.0cm}

  \vspace*{-2.8cm}\hspace*{7.5cm}{#2}\vspace*{2.8cm}
  \vspace*{0.5ex}
}

\newcommand{\eincardibild}[2]{
  \vspace*{1.5ex}
  \PSImagx{cardi_wav_#1.ps}{6.0cm}

  \vspace*{-3.5cm}\hspace*{5.5cm}{#2}\vspace*{3.5cm}

  \vspace*{0.5ex}
}

\BILD{tbh}
     {
      \hspace*{0.4cm}
      \begin{minipage}{8.5cm}
      \einstadbild{1992}{a)}

      \einstadbild{1660}{b)}

      \einstadbild{1771}{c)}
      \end{minipage}
      \hspace*{1cm}
      \begin{minipage}{8.5cm}
        \eincardibild{1816}{d)} 

        \vspace*{0.5cm}

        \eincardibild{1817}{e)}  
      \end{minipage}

     }
     {Left: Density plots $|\psi_n(q)|^2$ for 
      three different odd-odd eigenfunctions of the $a=1.8$ 
      stadium billiard:
      a) $n=1992$,  ``generic'' b) $n=1660$, bouncing ball mode
      c) $n=1771$ localized eigenfunction.
      Right: Density plots for two eigenfunctions of
      the cardioid billiard with odd symmetry: d) $n=1816$, ``generic''
      e) $n=1817$, localized along the $\overline{AB}$ orbit.
      Notice that according to the quantum ergodicity
      theorem the non-localized eigenfunctions 
      of type a) and d) are  the overwhelming majority.
     }
     {fig:stad-waves}

The third system is the cardioid billiard, which
is the limiting case of a family of billiards introduced in \cite{Rob83}.
The cardioid billiard is proven to be ergodic, mixing, a $K$-system
and a Bernoulli system \cite{Woj86,Sza92,Mar93,LivWoj95,CheHas96}.
Both the classical system \cite{Rob83,BruWhe96,BaeDul97,BaeChe97:p} 
and the quantum mechanical system 
have been studied
in detail \cite{Rob84,BaeSteSti95,BruWhe96,AurBaeSte97}.
The eigenvalues of the cardioid billiard have been provided
by Prosen and Robnik \cite{PrivComProRob} and were calculated
by means of the conformal mapping technique, see e.g.\ 
\cite{Rob84,BerRob86,ProRob93a}.
Using these eigenvalues, our study is based on computations for the first
6000 eigenfunctions of odd symmetry,
which were obtained from the eigenvalues
by means of the boundary integral method 
\cite{Rid79,BerWil84} using the singular value decomposition method
\cite{AurSte93}.
The boundary integral method was also used for the computations
of the eigenvalues and eigenfunctions of the stadium and the cosine billiard.

Let us first illustrate the structure of 
wave functions by showing density plots of $|\psi_n(q)|^2$ for 
three different types of wave functions of the stadium billiard 
and two different types of the cardioid billiard.
Fig.~\ref{fig:stad-waves}a) shows
a ``generic'' wave function,
whose density looks irregular.
Example b) belongs to the class of bouncing ball modes,
and its Wigner function is  localized in phase space on the 
bouncing ball orbits, see the discussion in section \ref{sec:Some-examples}. 
Fig.~\ref{fig:stad-waves}c) is another example of an eigenfunction showing
some kind of localization.
Fig.\ \ref{fig:stad-waves}d)
shows a ``generic'' wave function for the cardioid billiard and 
\ref{fig:stad-waves}e) is an example of an
eigenfunction, which shows a strong localization in
the surrounding of the shortest periodic orbit 
(with code $\overline{AB}$, see \cite{BruWhe96,BaeDul97}).

We should emphasize that according to the quantum ergodicity
theorem the overwhelming majority of states in the semiclassical limit
are of the type  a) and d),
which we also observe for the eigenfunctions of the studied
systems.

\newcommand{\einstadionplot}[2]{%
      \PSImagxy{#1.ps}{7.455cm}{3.3cm}

      \vspace*{-2.95cm}\hspace*{6.8cm}#2      

       \vspace*{2.95cm} 
      }

\newcommand{\eincosplot}[2]{%
      \PSImagxy{#1.ps}{7.9cm}{3cm}%
 
     \vspace*{-2.75cm}\hspace*{6.8cm}#2      

       \vspace*{2.75cm} 
     }

\newcommand{\eincardioidplot}[2]{%
      \PSImagxy{#1.ps}{6.7cm}{4.4cm}%

     \vspace*{-3.5cm}\hspace*{6.0cm}#2      

      \vspace*{3.5cm} 
       }

\BILD{b}
     {
      \hspace*{0.5cm}\begin{minipage}{8.5cm}
       \einstadionplot{stad_gebiete}{a)}
        
       \vspace*{2ex}

       \eincosplot{cosine_gebiete}{b)}
      \end{minipage}
     \begin{minipage}{7cm}
       \eincardioidplot{cardi_gebiete}{c)}
      \end{minipage}

       \vspace*{1ex}

     }
     {Shapes of the billiards studied numerically 
      in this work: a) desymmetrized
     stadium billiard, b) desymmetrized cosine billiard and c)
     desymmetrized cardioid billiard. 
     The rectangles in the interior of the billiards mark
     the domains $D_i$ of integration for studying the rate of quantum
     ergodicity in configuration space.  }
     {fig:billiard-domains}

%%%%%%%%%%%%%%%%%%%%%%%%%%%%%%%%%%
\subsection{Quantum ergodicity in coordinate space}
%%%%%%%%%%%%%%%%%%%%%%%%%%%%%%%%%%

\BILD{!ht}
     {
     \vspace*{-1.0cm}
     \begin{center}
                 \PSImagxy{stad_area_4.ps}{16cm}{10.cm}\hspace*{3cm} 

                 \vspace*{0.25cm}

                 \PSImagxy{odd_n__area_5.ps}{16cm}{10.0cm}\hspace*{3cm} 

                 \vspace*{-0.25cm}
     \end{center}

     }
     {Plot of $d_i(n) = 
      \int_{D_i} |\psi_{n}(q)|^2 \;\ud q- \tfrac{\Vol{(D_i)}}{\Vol{(\Omega)}}$ 
      for domain $4$ in the stadium billiard
      and for domain $5$ in the cardioid billiard. Since 
      $|\psi_{n}(q)|^2\geq 0$ one has $d_i(n)\geq- 
       \tfrac{\Vol{(D_i)}}{\Vol{(\Omega)}} $. For domain $D_4$ in the stadium
       this lower bound is attained by the bouncing ball modes whose 
      probability density $|\psi_{n}(q)|^2$ nearly vanishes in $D_4$; they 
      are responsible for the sharp edge seen in the plot of $d_4(n)$.
     }
     {fig:stad-qerg-area}

The quantum ergodicity theorem applied to the observable with symbol
$a(q)=\chi_D(q)$, discussed in section \ref{sec:Some-examples}, 
states that the difference
\begin{equation}
  \label{eq:an-diff}
  d_i(n) =
   \Int_{D_i} |\psi_{n}(q)|^2 \;\ud q- \frac{\Vol{(D_i)}}{\Vol{(\Omega)}} 
\end{equation}
vanishes for a subsequence of density one.
The first set of domains $D_i$ for which we investigate
the approach to the ergodic limit is shown 
in \figref{fig:billiard-domains}.
Plots of $d_i(n)$ for domain $D_4$ of the stadium billiard and $D_5$ 
of the cardioid billiard in \figref{fig:stad-qerg-area} show
quite large fluctuations around zero.
In particular for the stadium billiard there are many states
for which $d_1(n)$ is quite large and 
$d_4(n)$ is quite small. As one would expect, a large number 
of them are bouncing ball modes.
The fluctuations of $d_i(n)$ for the 
cosine billiard behave similarly to the stadium billiard.

When trying to study the rate of the approach to the
quantum-ergodic limit numerically
one therefore is faced with two problems.
On the one hand $d_i(n)$ is strongly fluctuating,
which makes an estimate of the approach to the mean
very difficult, if not impossible for the available numerical data.
On the other hand one does not know a priori which subsequences
should be excluded in \eqref{eq:an-diff}.
Therefore the investigation of the asymptotic behavior of the ``cumulative''
version \eqref{eqn:qet-sum-version} of the quantum ergodicity theorem
is much more appropriate. For the observable $\chi_D (q)$ we have 
\begin{equation}
S_1(E, \chi_{D})=\frac{1}{N(E)} \sum_{E_n \le E} 
\left|\langle \psi_n ,\chi_{D}\psi_n\rangle 
              -\frac{\Vol{(D)}}{\Vol{(\Omega)}}
\right|\,\, .
\end{equation}

In figs.\ \ref{fig:cos-qerg-S1}, \ref{fig:stad-qerg-S1} and 
\ref{fig:cardi-qerg-S1} we display 
$S_1(E,{\chi_{D_i}})$ for 
the different domains $D_i$, shown in fig.\ \ref{fig:billiard-domains},
in the desymmetrized cosine, stadium and cardioid
billiard, respectively. 
One nicely sees that the numerically determined curves for 
$S_1(E,{\chi_{D_i}})$ decrease with increasing energy.
This is of course expected from the quantum-ergodic
theorem, however 
since this is an asymptotic statement,
it is not clear a priori, whether one can observe
such a behavior also at low energies.
It should be emphasized that \figref{fig:cos-qerg-S1} is
based on the expectation values
$\langle \psi_n ,{\chi_{D_i}}  \psi_n  \rangle$ for 2000
eigenfunctions and  figs.~\ref{fig:stad-qerg-S1}
and \ref{fig:cardi-qerg-S1} are based on 6000 eigenfunctions in each case.

In order to study the rate of quantum ergodicity
quantitatively a fit of the function 
\begin{equation}
  \label{eq:S1-fit-fct}
  S_1^{\text{fit}} (E) = \ALPHA E^{-1/4+\varepsilon}
\end{equation}
to the numerical data for $S_1(E,{\chi_{D_i}})$ is performed.
As discussed in section \ref{sec:rate-of-quantum-ergodicity}, 
for certain systems a behavior
$S_1(E,\OpA)=O(E^{-1/4+\varepsilon})$ for all $\varepsilon>0$
is expected, so that the fit parameter
$\varepsilon$ characterizes the rate of quantum ergodicity.
A positive value of $\varepsilon$ thus means a slower decrease
of $S_1(E,\OpA)$ than the expected $E^{-1/4}$.
The results for $\varepsilon$ are shown in tables 
\ref{tab:qerg-S1-cos}--\ref{tab:qerg-S1-cardi},
and the insets in figs.\ \ref{fig:cos-qerg-S1}--\ref{fig:cardi-qerg-S1}
show the same curves
$S_1(E,\chi_{D_i})$ in a double--logarithmic plot together
with these fit curves.
The agreement of the fits with the computed functions $S_1(E,{\chi_{D_i}})$
is very good.
However, $\varepsilon$ is not small for all domains $D_i$
of the considered systems, rather we find several significant
exceptions, which will be explained in the following discussion.

%%%%%%%%%%%%%%%%%%%%%%%%%%%%%%%%%%%
\subsubsection{Cosine billiard} \label{sec:qerg-rate-cos-billiard}
%%%%%%%%%%%%%%%%%%%%%%%%%%%%%%%%%%%

For the cosine billiard one would expect a strong influence 
of the bouncing ball modes on the rate, since their 
number increases according to \cite{BaeSchSti97a} as
$N_{\text{bb}}(E)\sim c\;  E^{9/10}$. 
But the prefactor 
$c$ turns out to be very small and therefore 
the influence of the bouncing ball 
modes is suppressed at low energies. 
The model for $S_1(E,A)$, equation (\ref{mod-rate}), 
gives for the cosine billiard
\begin{equation}\label{s1-cosinus}
S_1^{\text{model}} (E,\chi_{D_i})=(1-0.201\, E^{-0.13})\, \slim'(\chi_{D_i})E^{-\frac{1}{4}}
+0.201\,  \slimbb''(\chi_{D_i})E^{-0.13}\,\, ,
\end{equation}
where we have inserted the values $c=0.04$ and $\beta =0.87$, obtained 
in \cite{BaeSchSti97a} from a fit to $N_{\text{bb}}(E)$ which was 
performed over the 
same energy range which we consider here. For sake of completeness 
we have included the first factor, $(1-0.201\, E^{-0.13})$, but the 
numerical fits we perform below only change marginally if one sets 
this factor equal to 1. 

\BILD{tbh}
     {
        \PSImagxy{cosine_area_kumulativ1.ps}{16cm}{11cm}\hspace*{1cm}
     }
     {Plot of $S_1(E,{\chi_{D_i}})$ for different domains $D_i$ 
              for the cosine billiard using the first 2000 eigenfunctions,
      see fig.~\ref{fig:billiard-domains}b) for the location 
      of the domains $D_i$.
      The inset shows the same curves in double--logarithmic representation 
      together with a fit of 
      $S_1^{\text{fit}}(E) = \ALPHA E^{-1/4+\varepsilon}$ 
      to the numerical data.}
     {fig:cos-qerg-S1}

\begin{table}[!t]
  \vspace*{3.5ex}
  \begin{center}
   \renewcommand{\arraystretch}{1.25}
  \begin{tabular}{|c|c||c||c|c|c|}\hline
    domain & rel. area & $ \varepsilon$ & $\ALPHA$ & $\slim'(\chi_{D_i})$  &  $\slimbb''(\chi_{D_i})$ \\ \hline
    1      & 0.018    & $ -0.002$   &  0.052    &   0.0525     &  0.0045       \\          
    2      & 0.018    & $ +0.012$   &  0.026    &   0.0468     &  0.0067       \\          
    3      & 0.008    & $ +0.013$   &  0.043    &   0.0297     &  0.0020       \\           
    4      & 0.008    & $ +0.022$   &  0.023    &   0.0273     &  0.0030       \\            
    5      & 0.015    & $ +0.020$   &  0.050    &   0.0543     &  0.0150       \\ \hline   
    6      & 0.336    & $ +0.009$   &  0.258    &   0.2471     &  0.0840       \\          
    7      & 0.512    & $ +0.023$   &  0.352    &   0.2920     &  0.1280       \\           
    8      & 0.648    & $ +0.009$   &  0.381    &   0.3410     &  0.1620       \\            
    9      & 0.800    & $ +0.054$   &  0.279    &   0.3264     &  0.2500       \\ \hline   
\end{tabular}
\end{center}
  \Caption{ Rate of quantum ergodicity for the cosine billiard
           with domains $D_i$ as 
           shown in fig.~\ref{fig:billiard-domains} and 
           fig.~\ref{fig:cos-qerg-S1} and in the inset of 
           fig.~\ref{fig:cos-qerg-S1-zusatz1}.
           Shown are the results for $\varepsilon$ and $\ALPHA$ of the fit
           of $S_1^{\text{fit}}(E) = \ALPHA E^{-1/4+\varepsilon}$ 
           to the numerical data.
           Also tabulated are the values for the relative area
           of the corresponding domains,
           the quantities $\slimbb''(\chi_{D_i})$ computed
           according to \eqref{eq:qlim-bb-cosine}
           and the result $\slim'(\chi_{D_i})$ of the
           fit of the model \eqref{s1-cosinus} to $S_1(E,{\chi_{D_i}})$.
           }{tab:qerg-S1-cos}
\end{table}

The asymptotic behavior of the probability density $|\psi_{n''}(q)|^2$ of the 
bouncing ball modes is (in the weak sense)
\begin{equation}
|\psi_{n''}(q)|^2\sim
\begin{cases} 1/\Vol (R) & \text{for}\,\, q\in R                     \\
                 0     & \text{for}\,\, q\in \Omega\backslash R
\end{cases}\,\, ,\qquad \text{as}\,\, n'' \to\infty \,\, ,
\end{equation}
 where $R$ denotes the rectangular part 
of the billiard.  
So the expectation values are asymptotically 
$\langle\psi_{n''}, \chi_D \psi_{n''}\rangle \sim \Vol (D\cap R)/\Vol (R)$, 
and since
$\slimbb ''(\chi_{D})=\lim_{E\to\infty}S''(E,\chi_{D})$ is the mean value 
of $|\langle\psi_{n''}, \chi_D \psi_{n''}\rangle -\Vol (D)/\Vol(\Omega) |$
over all bouncing ball modes one has 
\begin{equation} \label{eq:qlim-bb-cosine}
\slimbb ''(\chi_{D})=
\left| \frac{\Vol (D\cap R)}{\Vol (R)}-
\frac{\Vol (D)}{\Vol (\Omega)} \right|\,\, . 
\end{equation}
For fixed volume $\Vol (D)$ the quantity 
$\slimbb ''(\chi_{D})$ is maximal for domains $D$ lying 
entirely outside of the rectangular region, 
$\slimbb ''(\chi_{D}) =\tfrac{\Vol (D)}{\Vol (\Omega)} $.
For domains lying entirely inside the rectangular part of the billiard,
we have the minimal value 
$\slimbb ''(\chi_{D}) =\tfrac{1}{4}\tfrac{\Vol (D)}{\Vol (\Omega)}$.
Therefore the strongest contribution of the bouncing ball modes 
to $S_1(E,\chi_D)$ in eq.~(\ref{s1-cosinus}) is expected for the domains  
outside the rectangular region.

The values for $\slimbb ''(\chi_{D_i})$ are given in 
\tabref{tab:qerg-S1-cos}. The largest values for the small 
domains are obtained for the 
domains outside the rectangular part of the billiard for which also 
the rate of quantum ergodicity is the slowest.
Furthermore we 
see from \tabref{tab:qerg-S1-cos} that the 
factor $0.201\,  \slimbb''(\chi_{D_i})$ in front of 
$E^{-0.13}$ in equation (\ref{s1-cosinus}) is for all domains 
much smaller than the prefactor $\ALPHA$ from the fit to 
(\ref{eq:S1-fit-fct}). This  already indicates that the contribution 
of the bouncing ball modes is suppressed, explaining 
why the rate for the cosine billiard is in such a good agreement with $\varepsilon=0$.

\BILD{tbh}
     {
        \PSImagxy{cosine_area_kumulativ1_zusatz1.ps}{16cm}{10.0cm}\hspace*{1cm}
     }
     {Plot of $S_1(E,{\chi_{D_i}})$ for two further domains $D_8$  and $D_9$
         (dashed curve) in the cosine billiard using the first
      2000 eigenfunctions.
      Also shown is the fit
      $S_1^{\text{model}}(E,\chi_{D_i})$, eq.~\eqref{s1-cosinus}.
      }
     {fig:cos-qerg-S1-zusatz1}

In order to test this quantitatively
we have performed a fit of the model 
(\ref{s1-cosinus}) to the numerical data, where the only 
free parameter is $\slim '(\chi_{D_i})$. The accuracy of the fits 
is very good and the results for $\slim '(\chi_{D_i})$ are 
shown in \tabref{tab:qerg-S1-cos};  they are much larger than the 
corresponding prefactors $0.201\,  \slimbb''(\chi_{D_i})$ of the 
bouncing ball part of $S_1(E,\chi_{D_i})$. 
Therefore the influence of the bouncing ball modes on the rate 
is negligible small on the present energy interval, despite the fact 
that asymptotically they should dominate the rate. 

The domains $D_3\subset D_1$ and $D_4 \subset D_2$
show a slightly slower rate than $D_1$ and $D_2$, respectively.
This is due to the fact that choosing a smaller domain $D$
implies larger fluctuations of  
$\langle \psi_n , {\chi_{D}} \psi_n \rangle$
for the same set of eigenfunctions.

As an additional test we have computed $S_1(E,\chi_{D_i})$ numerically 
for four further domains (shown in the inset 
of fig.~\ref{fig:cos-qerg-S1-zusatz1}), 
having a much larger area than the previous 
ones. For these 
domains  $\slimbb ''(\chi_{D_i})$ is larger, and one therefore expects 
a stronger influence of the bouncing ball modes and correspondingly 
a slower rate of quantum ergodicity. The results are shown in 
table \ref{tab:qerg-S1-cos} and fig.~\ref{fig:cos-qerg-S1-zusatz1}
and our findings are completely consistent with the previous 
ones as well as with the model  (\ref{s1-cosinus}). 
We also observe in \figref{fig:cos-qerg-S1-zusatz1} 
that for the large domains, except 
for the whole rectangular part $D_9=R$, the rate is faster 
at low energies, than at high energies.
This is due to the influence of the boundary and will 
be discussed in section \ref{sec:boundary-effects}.

Summarizing the results for the cosine billiard, we found that
the rate of quantum ergodicity 
is in excellent  agreement with a rate
proportional to $E^{-1/4}$ for the subsequence of
quantum-ergodic eigenfunctions. The phenomenological model 
$S_1^{\text{model}}(E,\chi_{D})$, eq.\ (\ref{s1-cosinus}), is 
in very good agreement with the numerical data, especially in view 
of the fact that it contains only one free parameter.  
Furthermore, the cosine billiard provides an impressive example 
of a system for which the asymptotic regime for $S_1(E,A)$ 
is reached very late. Up to the 2000th eigenfunction the 
asymptotic behavior $S_1(E,A)\sim C E^{-1/10}$ is almost 
completely hidden. A continuation of $S_1^{\text{model}}(E,\chi_{R})$
for the  domain $R=D_9$ with the strongest 
influence of the bouncing ball modes, shows that 
at $E\approx 10^6$ the two contributions have the same magnitude, and 
one has to go  up as high as $E\approx 10^{20}$ to see the asymptotic behavior
$S_1(E,\chi_R )\sim C E^{-1/10}$.  
Therefore there is no contradiction between the observed fast 
rate of quantum ergodicity in the present energy range
and the increase of the number of bouncing ball modes 
$N_{\text{bb}}(E)\sim c\;  E^{9/10}$ found in \cite{BaeSchSti97a}.

%%%%%%%%%%%%%%%%%%%%%%%%%%%%%%%%%%%%%%%%%%%%%%
\subsubsection{Stadium billiard}
%%%%%%%%%%%%%%%%%%%%%%%%%%%%%%%%%%%%%%%%%%%%%%

For the stadium billiard the number of bouncing 
ball modes grows as $N_{\text{bb}}(E)\sim c\,  E^{3/4}$  
\cite{Tan97,BaeSchSti97a}. 
Therefore the bouncing ball mode contribution 
to $S_1(E,A)$ is, according to equation (\ref{mod-rate}), 
proportional to $E^{-1/4}$, and thus of the same order as the 
expected rate of quantum ergodicity for the quantum-ergodic 
eigenfunctions. One therefore expects for all domains 
in position space a rate of $E^{-1/4}$. 
We have investigated the rate of quantum ergodicity for
the stadium billiard using the small domains shown in 
fig.~\ref{fig:billiard-domains}a)
and for  larger domains shown in fig.~\ref{fig:stadium-domains-extra}.
The results of the fits of 
$S_1^{\text{fit}} (E) = \ALPHA E^{-1/4+\varepsilon}$ to 
the numerical data for $S_1(E,\chi_{D_i})$ 
are given in table \ref{tab:qerg-S1-stad}.

Let us first discuss the rate for the small domains shown in 
fig.~\ref{fig:billiard-domains}a).
For the domains $D_1$ and $D_2$ which lie inside the rectangular 
part of the billiard the rate is in very good agreement 
with $E^{-1/4}$. But both for the domain $D_3$ which lies 
on the border between the rectangular part and the quarter circle,  
and in particular for domain $D_4$ 
which lies inside the quarter circle, one finds a slower rate
than expected. This is a behavior which 
one would expect for a billiard with a much faster increasing 
number of bouncing ball modes.  

\BILD{tbh}
     {
  \vspace*{-3.5ex}
       \PSImagxy{stad_area_kumulativ1.ps}{16cm}{11.0cm}\hspace*{1cm}
     }
     {Plot of $S_1(E,{\chi_{D_i}})$ for different domains $D_i$ for
      the stadium billiard using the first 6000 eigenfunctions,
      see fig.~\ref{fig:billiard-domains}a) for the location 
      of the domains $D_i$. The inset
      shows the same curves in double--logarithmic representation 
      together with a fit of eq.~\eqref{eq:S1-fit-fct}.}
     {fig:stad-qerg-S1}

\begin{table}[b]
%  \vspace*{3.5ex}
  \begin{center}
   \renewcommand{\arraystretch}{1.25}
  \begin{tabular}{|c|c||c||c|c|c|}\hline
    domain & rel. area & $ \varepsilon$ & $\ALPHA$ & $\slim'(\chi_{D_i})$ & $b(A)$  \\ \hline
    1      &  0.015  & $ + 0.009$  & 0.041   &  0.0539  &  0.0000  \\          
    2      &  0.015  & $ + 0.012$  & 0.041   &  0.0564  &  0.0000  \\          
    3      &  0.015  & $ + 0.033$  & 0.035   &  0.0533  &  0.0008  \\           
    4      &  0.015  & $ + 0.095$  & 0.029   &  0.0492  &  0.0047  \\\hline            
5  &  0.015  & $ + 0.020$  & 0.039   &  0.0551  &  0.0004  \\      % fit ab 600:     
6   &  0.278  & $ + 0.070$  & 0.137   &  0.1401  &  0.0233  \\            
7   &  0.433  & $ + 0.111$  & 0.118   &  0.1071  &  0.0395  \\           
8   &  0.557  & $ + 0.168$  & 0.089   &  0.0292  &  0.0634  \\            
9   &  0.696  & $ + 0.188$  & 0.098   &  0.0384  &  0.0827  \\            
10  &  0.681  & $ + 0.084$  & 0.176   &  0.2474  &  0.0295  \\ \hline   
\end{tabular}
\end{center}
  \Caption{Rate of quantum ergodicity for the stadium billiard with 
           domains $D_i$ as 
           shown in figs.~\ref{fig:billiard-domains} and 
           \ref{fig:stadium-domains-extra}.
           Shown are the results for $\varepsilon$ and $\ALPHA$ 
           of the fit
           $S_1^{\text{fit}}(E) = \ALPHA E^{-1/4+\varepsilon}$ to 
           $S_1(E,{\chi_{D_i}})$.
           Also tabulated are the values for the relative area
           of the corresponding domains,
           and the results $\slim'(\chi_{D_i})$ and $b(A)$
           of the fit of the model \eqref{s1-stadium2} to 
           $S_1(E,{\chi_{D_i}})$.
          }{tab:qerg-S1-stad}
\end{table}

%    1      &  0.01547  & $ + 0.009$  & 0.0405   &  0.034     & 0.0005  \\          
%    2      &  0.01547  & $ + 0.012$  & 0.0409   &  0.035     & 0.0007  \\          
%    3      &  0.01547  & $ + 0.033$  & 0.0353   &  0.032     & 0.0015  \\           
%    4      &  0.01547  & $ + 0.095$  & 0.0288   &  0.030     & 0.0053  \\\hline            
%5  (typ100)&  0.01547  & $ + 0.020$  & 0.0387   &  0.032     & 0.0013  \\      % fit ab 600:     
%6  (typ10) &  0.27849  & $ + 0.070$  & 0.1372   &  0.084     & 0.0253  \\            
%7  (typ1)  &  0.43320  & $ + 0.111$  & 0.1182   &  0.064     & 0.0410  \\           
%8  (typ7)  &  0.55697  & $ + 0.168$  & 0.0888   &  0.017     & 0.0638  \\            
%9  (typ2r) &  0.69622  & $ + 0.188$  & 0.0982   &  0.023     & 0.0833  \\            
%10 (typ50) &  0.68075  & $ + 0.084$  & 0.1764   &  0.148     & 0.0330  \\ \hline   

We see three possible explanations for this behavior of the rate 
for the stadium billiard. First, the counting function 
$N_{\text{bb}}(E)$ for the bouncing ball modes might increase with a larger 
exponent than $3/4$, $N_{\text{bb}}(E)\sim c\, E^{\beta}$, $\beta >3/4$. 
This would contradict the results in \cite{Tan97,BaeSchSti97a}, derived by 
independent  methods. Moreover, the 
exponent $\beta$ was tested numerically in \cite{BaeSchSti97a} up to energy 
$E\approx 10 000$ and we found 
very good agreement with $\beta=3/4$. Even if we relaxed the 
criteria for the selection of the bouncing ball modes drastically, 
the exponent did not change significantly, only the prefactor $c$ increased. 
Therefore we think that this first possibility is clearly ruled out.

Secondly, the rate for the quantum-ergodic eigenfunctions might 
not be proportional to $E^{-1/4}$, but has a slower decay rate. 
Then we have to 
assume a position dependence of the rate, in order to explain the 
different behavior for the different domains: in the rectangular 
part of the billiard the rate has to be proportional to $E^{-1/4}$ to 
explain the value of $\varepsilon$ obtained for the domains $D_1$ and $D_2$. 
 Whereas 
inside the quarter circle the rate of decay has to decrease as 
$S_1'(E,\chi_{D_4})\sim \slim '(A) E^{-0.15}$, in order to explain 
the value of $\varepsilon$ obtained for  $D_3$ and $D_4$. 
A priori such a dependence of the 
rate of the quantum-ergodic eigenfunctions on 
the location of the domain in the billiard 
is not impossible.
If this is the case 
then one should observe no dependence of the rate 
on the volume of the domain $D$, as long as one stays in 
the same region of the billiard. E.g.\ the rate for a domain like 
$D_6$, 
which contains $D_1$ and $D_2$ and is far enough away from 
the quarter circle, should be the same as the one for 
$D_1$ and $D_2$.

The third possible explanation for the observed behavior of the 
rate is that there exist more not quantum-ergodic eigenfunctions 
which have a larger probability density in the rectangular 
part than in the quarter-circle, and which are not bouncing ball 
modes. 
Alternatively the reason could be a subsequence of density zero
of quantum-ergodic eigenfunctions, which has 
a sufficiently increasing counting function
and a slow rate, see the remark at the end of section 
\ref{sec:rate-of-quantum-ergodicity}. 
In both cases 
the model for $S_1(E,A)$ discussed in section 
\ref{sec:rate-of-quantum-ergodicity}, which we 
already used in the case of 
the cosine billiard, would be applicable. 
In contrast to the second possibility   
in this scenario one expects a dependence of the rate of $S_1(E,\chi_D)$ on 
the volume of the domain $D$, as in the case of
the cosine billiard.

\BILD
     {!t}
     {
       \begin{center}
           \einstadionplot{stad_gebiete_extra}{}
             
           \vspace*{-4ex}
       \end{center}
     }
     {Domains in the $a=1.8$ stadium billiard used to decide between
      the different explanations for the slow rates in the stadium billiard.}
     {fig:stadium-domains-extra}

To decide which explanation is the correct one we 
studied the rate for a number of large domains 
shown in fig.~\ref{fig:stadium-domains-extra}. With these domains one 
necessarily comes closer to the boundary of the billiard. 
To rule out the possibility that the observed behavior of 
the rate is due to the influence of the boundary, and not due
to the dependence on the volume and location 
of the domains, we  computed in addition $S_1(E,\chi_D)$ 
for the small domain $D_5$ which is close to the boundary.    

The results are also given in table \ref{tab:qerg-S1-stad} and
some examples of $S_1(E,\chi_{D_i})$ for these large domains are
shown in fig.~\ref{fig:stad-qerg-S1-large-domain}.
As for the cosine billiard, we also found that for large domains 
at small energies the
rate may be much faster than at higher energies
which is nicely seen in fig.~\ref{fig:stad-qerg-S1-large-domain}
for the domains $D_7$ and $D_8$.
This effect is due to the influence of the boundary, as we 
will  discuss 
in section \ref{sec:boundary-effects}; here we only note that 
the boundary influence vanishes for large energies.
 
The observed rate  of quantum ergodicity displays a strong 
dependence on the volume of the domain $D$, whereas the 
location, as long as one stays inside the rectangular part, 
has no influence. E.g.\ for the domain $D_6$, which 
contains $D_1$ and $D_2$, one gets a much slower rate than for 
$D_1$ and $D_2$. In contrast to $D_6$ the rate for the small 
domain $D_5$ near the boundary is rather close to the one 
for $D_1$ and $D_2$. The slightly slower rate for $D_5$ is 
due to the smaller energy range for which we have computed 
$S_1(E,\chi_{D_5})$. A fit of 
$S_1^{\text{fit}} (E) = \ALPHA E^{-1/4+\varepsilon}$ 
to $S_1(E,\chi_{D_1})$ and $S_1(E,\chi_{D_2})$ 
using the first 2000 eigenfunctions gives an $\varepsilon$ of 
$0.022$ for $D_1$ and $0.011$ for $D_2$, which is of the same 
magnitude as the result for $D_5$. 
Moreover the rate decreases monotonically with increasing area 
of the domains $D_i$, as long as they are inside the rectangular 
part $R$ of the billiard.

The domain $D_{10}$ is interesting because it  extends over both 
parts of the billiard. The enhanced probability 
density  of the exceptional eigenfunctions in the rectangular part 
is partially compensated by the lower probability density in 
the quarter circle. Therefore one expects a rate similar 
to a domain in the rectangular part with relative area 
$(\Vol (D_{10})-2\Vol (D_{10}\cap (\Omega\backslash R)))/\Vol (\Omega )
=0.371\ldots $. This relative area lies between the values for 
$D_6$ and $D_7$, and 
indeed the rate for $D_{10}$ lies 
between the rate for $D_6$ and $D_7$ too.

These results strongly support the third explanation, i.e.\ the 
existence of a large density zero   subsequence which is 
responsible for the deviations of the rate from $E^{-1/4}$. 
``Large'' means that the counting function increases
sufficiently strong to cause the rate to deviate from 
the expected behavior.

To test this conjecture quantitatively one has to compare the numerical 
data with the  conjectured behavior 
\begin{equation}\label{s1-stadium}
S_1^{\text{model}} (E,A)=\left(1-c E^{-\beta}\right)
\slim '(A)E^{-1/4}+b(A) E^{-\beta}\,\, .
\end{equation}
Since this model contains the four free parameters
$c$, $\beta$, $\slim '(A)$ and $b(A)$, the numerical 
fit is not very stable.
Therefore it is desirable to get some 
additional information from a different source. 

\BILD
     {t}
     {
       \PSImagxy{stad_areagrosses_recht_typ7_kumulativ.ps}{16cm}{10.5cm}\hspace*{1cm}
     }
     {Plot of $S_1(E,{\chi_{D}})$ for large domains (see fig.~\ref{fig:stadium-domains-extra})
              for the $a=1.8$ stadium billiard using the first 
              2000 eigenfunctions. The inset
      shows the same curves in double--logarithmic representation 
      together with a fit of eq.~\eqref{eq:S1-fit-fct}.
              For the domains $D_7$ and in particular for domain $D_8$
              a sharp transition from a fast to a slower decay of the
              rate is visible. This effect is due to the boundary and will be 
              explained in sec.~\ref{sec:boundary-effects}.
}
     {fig:stad-qerg-S1-large-domain}

To this end we plotted  
$d_4(n)=\langle \psi_n ,\chi_{D_4}\psi_n\rangle -\Vol (D_4)/\Vol(\Omega) $ 
for domain $D_4$ which shows a 
slow rate, see \figref{fig:stad-qerg-area},  
 and divided the spectrum into two parts by inserting 
a curve $-c_{\text{d}} E^{-1/4}$, and a curve 
$c_{\text{u}} E^{-1/4}$. The part of the spectrum between 
the two curves corresponds 
to the quantum-ergodic eigenfunctions 
with the optimal rate $\sim E^{-1/4}$, and the part above and 
below the curves corresponds to the not quantum-ergodic eigenfunctions or 
to quantum-ergodic eigenfunctions with a slower rate than 
$E^{-1/4}$. By computing the counting functions for these two subsequences 
we get a further criterion for 
distinguishing between the two possible scenarios for 
the behavior of the eigenfunctions discussed above. 
If the rate of quantum ergodicity for all quantum-ergodic 
eigenfunctions is slower than $E^{-1/4}$ inside the quarter 
circle, then the fraction of eigenfunctions which lie 
below or above the two curves should grow proportional to $E$. 
If the deviation of the rate is due to a not quantum-ergodic 
subsequence of density zero, or a quantum-ergodic 
subsequence of density zero with 
exceptionally slow rate,  then the number of 
states which lie 
below or above the two curves should grow like $N''(E)$, i.e.\
slower than $E$.

Proceeding in the described way, we find that 
the majority of the exceptional states
have values of $d_4(n)$ which 
lies below the lower curve $-c_{\text{d}} E^{-1/4}$.
A numerical fit for their counting function 
gives $N'' (E)=0.06 \,E^{0.93}$. The exponent is very stable under 
slight variations of the constant $c_{\text{d}}$
which determines the curve. Up to $E\approx 10000$ corresponding to 
the 2000th state, the counting function even has an almost linear
behavior.
The nature of these states will be discussed below.

The numerical result $N'' (E)=0.06 \,E^{0.93}$ allows to determine the 
parameters $c=0.06\frac{ 4\pi}{\Vol (\Omega )}= 0.29\ldots$ 
and $\beta = -0.07$ in 
the model (\ref{s1-stadium}) giving
\begin{equation}\label{s1-stadium2}
S_1^{\text{model}}(E,A)=
\left(1-0.29\, E^{-0.07}\right) \slim ' (A) E^{-1/4}+b(A) E^{-0.07}\,\, .
\end{equation}
We have now eliminated two of the four free parameters, and can 
therefore test this formula effectively with the numerical data. 
The results for $\slim' (\chi_{D_i})$ and $b (\chi_{D_i})$ are also shown in 
table \ref{tab:qerg-S1-stad}, and for three large domains the plot of 
$S_1(E,\chi_D)$ and the corresponding fit $S_1^{\text{model}}(E,\chi_D)$
is shown in \figref{fig:stad-qerg-S1-large-domain}. 

The agreement of the fits with the numerical data is very good. 
Moreover the values for $\slim '(D_i)$ and $b(D_i)$ are very reasonable:
For $\slim '(D_i)$ one expects that it depends on 
the volume of $D_i$ only, and not on the location. This is very well 
confirmed for the domains $D_1$--$D_5$, which have the same volume, where 
$\slim '(D_i)$ stays almost constant. 
According to the results in 
\cite{AurTag97:p} one expects that $\slim '(D_i)$ increases with 
increasing volume of $D_i$, for small $\Vol (D_i)$, 
then reaches a plateau, and then finally decreases for very large domains. 
This behavior is not observed, 
the values for $\slim '(D_i)$ rather oscillate.
The most striking difference occurs between   $\slim ' (D_9)$ and 
 $\slim ' (D_{10})$, because they have approximately the same volume. 
We furthermore find
that the behavior of $\slim '(D_i)$ 
is completely analogous to that of $\ALPHA_i$. 
The  behavior of $b(D_i)$ is in perfect accordance with what one 
expects for a sum of quantum limits which are concentrated 
on the rectangular part of the billiard. The values increase when 
moving $D_i$ into the quarter circle, and they increase 
with increasing volume of  $D_i$, as long as $D_i$ lies entirely  inside the 
rectangular part. For $D_{10}$ the parameter $b(D_{10})$ takes an 
intermediate value between $b(D_{6})$ and $b(D_{7})$, as one expects.

The inclusion of the factor $(1-0.29\, E^{-0.07})$ in 
eq.~(\ref{s1-stadium2}) turned out to be necessary to get 
satisfactory results. The contribution
of $E^{-0.07}$ cannot be neglected in the present energy range
because of the small exponent.
Without this factor we obtained 
for some of the domains negative values for $\slim ' (D_i)$, 
which is impossible because $S_1'(E,A)$ is by definition 
positive. 

\BILD{!b}
     {
       \PSImagxy{raten_vergleich.ps}{16cm}{11cm}\hspace*{1cm} 
     }
     {Plot of $S_2(E,\chi_{D_i})$ for the domains $D_1$ and  $D_2$ 
      in the stadium billiard. The dashed lines show the fit of the 
      conjectured behavior  
      $ c\, E^{-1/2}
       \ln  \bigl(\tfrac{\Vol(\Omega)}{2} E^{1/2}\bigr)$ to  
       $S_2(E,\chi_{D_i})$.
      %One sees that the agreement of 
      %the fits with the numerical data is very poor.}
      The result of the fit shows that the numerical
      data for the first 6000 expectation values
      cannot be described with this rate.}
     {fig:vergleich}

This also sheds some light on the limitations of such a simple model 
like (\ref{s1-stadium2}). For the exponent $\beta=0.07$ only the order 
of magnitude is known for sure, the constant $c$ from $N'' (E)$ 
might still vary, and nothing is known about the behavior
of the higher order contributions to 
$S_1'$ and $S_1''$. 
In view of this it is 
surprising how good this model fits with the numerical data. 
We believe that this gives a very strong support 
for the underlying conjectures, namely that 
a density one subsequence of quantum-ergodic eigenfunctions 
has a rate $S_1'(E,A)\sim cE^{-1/4}$, and the deviations 
in the rate of $S_1(E,A)$ from this behavior are due 
to a subsequence of density zero.

As mentioned in section \ref{sec:rate-of-quantum-ergodicity}, a behavior 
$S_2(E,A)\sim cE^{-1/2}
\ln \bigl(\tfrac{\Vol(\Omega)}{2} E^{1/2}\bigr)$ 
for the stadium billiard is 
claimed in \cite{EckFisKeaAgaMaiMue95}. 
We have tested this both for the small domains $D_1$ and $D_2$, which are 
not influenced by the bouncing ball modes and also for
some larger domains. 
However, the resulting fits clearly show that this result does not
apply to our numerical data, see fig.~\ref{fig:vergleich}.
We also tested if this result applies to
the quantum-ergodic subsequence, 
i.e.\ $S_1'(E,A)\sim cE^{-1/4}
\sqrt{\ln \bigl(\tfrac{\Vol(\Omega)}{2} E^{1/2}\bigr)}$, 
by replacing the term $E^{-1/4}$ in 
equation (\ref{s1-stadium2}) by $E^{-1/4}\sqrt{ \ln \bigl(\tfrac{\Vol(\Omega)}{2} 
E^{1/2}\bigr)}$.
Again we find that from our numerical data that this 
possibility is excluded, at least 
 for the energy range under consideration.
For the stadium billiard it is known that 
the asymptotic behavior  of the classical autocorrelation 
$C(\tau )\sim 1/\tau$, which leads to 
$S_2(E,A)\sim cE^{-1/2}
\ln \bigl(\tfrac{\Vol(\Omega)}{2} E^{1/2}\bigr)$ according to 
\cite{EckFisKeaAgaMaiMue95}, sets in rather late.
So it would be 
very interesting to compare the results 
with those obtained by inserting the numerically computed 
autocorrelation function in the integral in \eqref{eq:gen-rate}.

\BILD{tb}
     {
      \begin{minipage}{8.5cm}
      \einstadbild{1643}{a)}

      \einstadbild{1797}{c)}

      \end{minipage}
      \hspace*{0.5cm}
      \begin{minipage}{8.5cm}
      \einstadbild{1652}{b)} %*

      \einstadbild{1834}{d)} %*

      \end{minipage}
     }
     {Four examples of the exceptional eigenfunctions showing
      localization in the rectangular part of the stadium billiard,
      which are not bouncing ball modes, 
      a) $n=1643$, b) $n=1652$, c) $n=1797$ and d) $n=1834$.
     }
     {fig:exceptional-non-bb-modes}

We now return 
to the question of what type these additional 
subsequences of eigenfunctions are.
As additional information for the model, the counting 
function for the number of states for which 
$\langle \psi_n , \chi_{D_4} \psi_n\rangle -\tfrac{\Vol(D_4)}{\Vol(\Omega)}$ 
is smaller than $-c_{\text{d}} E^{-1/4}$ has been used.
For comparison we have carried out the same procedure for the observable
$1-\chi_{D_9}$ which corresponds to the complete quarter circle.
As expected the bouncing ball modes appeared in 
both subsequences, but additionally a considerable number 
of other types of eigenfunctions showed up.
In \figref{fig:exceptional-non-bb-modes} we show some examples 
of such eigenfunctions. 
They all show a reduced probability density inside the quarter circle, but 
their structure is essentially different from the  bouncing ball modes. 
Their semiclassical origin are maybe 
periodic orbits bouncing up and down between the
two perpendicular walls for a long time but then leaving 
the neighborhood of the  bouncing
ball orbits in phase space. 
At least it seems difficult to
associate short unstable periodic orbits
to the patterns in the shown states, because the lines of enhanced
probability do not always obey the laws of reflection,  
or they look too irregular.

\begin{table}[b]
  \vspace*{3.5ex}
  \begin{center}
   \renewcommand{\arraystretch}{1.25}
  \begin{tabular}{|l|c|c|c|}\hline
system       & domain A & domain B & domain C  \\\hline
stadium $(a=0.5)$  & $ + 0.111$      
                   & $ + 0.062$   
                   & $ + 0.056$   
                    \\ \hline
stadium $(a=1.8)$  & $ + 0.009$   
                   & $ + 0.033$   
                   & $ + 0.095$   
                    \\ \hline
stadium $(a=4.0)$  & $ - 0.008$   
                   & $ + 0.031$   
                   & $ + 0.095$   
                    \\ \hline
\end{tabular}
\end{center}
  \Caption{Results for $\varepsilon$ of the fit of  
           $S_1^{\text{fit}}(E) = \ALPHA E^{-1/4+\varepsilon}$ to 
           the numerically obtained $S_1(E,{\chi_{D_i}})$,  
           for stadium billiards with different
           parameter $a$ for three different domains $A$, $B$ and $C$. 
           Domain $A$ lies within the rectangular 
           part of the billiard,
           domain $B$ is centered at $x=a$ and domain $C$ is located
           in the quarter circle.}
  {tab:fit-different-stadiums}
\end{table}

A further test of the hypothesis that a density zero subsequence 
is responsible for the slow rate is provided by varying the length 
$a$ of the billiard. 
Here we used the first 2000 eigenfunctions for both the $a=0.5$
and the $a=4.0$ stadium billiard
in addition to the results for the $a=1.8$ stadium based on 6000
eigenfunctions.
We have chosen three different domains for these
three systems: domain $A$ lies within the rectangular part of the billiard,
domain $B$ is centered at $x=a$ and domain $C$ is located
in the quarter circle.
The results for the rate of quantum ergodicity are shown in 
\tabref{tab:fit-different-stadiums}.
For different parameters the quantities $b (D_i)$ change, and therefore 
the weights of the different contributions to $S_1(E,A)$ in equation 
(\ref{s1-stadium2}). 
For smaller $a$ the relative fraction of the volume 
of the rectangular part, $\Vol (R)/\Vol (\Omega)$ becomes smaller.
Therefore one expects that for smaller $a$
the influence of the 
not quantum-ergodic subsequences to $S_1(E,\chi_D)$ becomes stronger in the 
rectangular part, and weaker in the quarter circle.
This is nicely seen in the numerically found behavior of the rate for the 
domains
$A$ and $C$ shown in table \ref{tab:fit-different-stadiums},
which confirms our hypothesis.

To summarize our results for the stadium billiard, we have shown 
the existence of a large, but density zero,  
subsequence of 
eigenfunctions which have an enhanced probability distribution 
on the rectangular part of the billiard but 
having a different structure than the bouncing ball modes.
We  demonstrated that the observed effects are due to the influence 
of this  subsequence of density zero.
This subsequence shows a different behavior 
than the majority of quantum-ergodic eigenfunctions for which our results
imply a uniform rate of $E^{-1/4}$. Clearly we cannot decide if this 
exceptional subsequence will ultimately be not quantum-ergodic, 
or if it is a quantum-ergodic subsequence with a exceptional 
behavior of the rate. We can only say that on the 
presently studied energy range up to 
$E\approx 30 000 $, i.e.\ up to the $6000$th eigenfunction, they 
behave not quantum-ergodic.

%%%%%%%%%%%%%%%%%%%%%%%%%%%%%%%%%%%%%
\subsubsection{Cardioid billiard}
%%%%%%%%%%%%%%%%%%%%%%%%%%%%%%%%%%%%%

The cardioid billiard is probably the most ``generic'' 
one of our three billiards, in the sense that it possesses no 
two dimensional family of periodic orbits like the bouncing ball orbits. 
One might therefore expect a priori 
a better rate of quantum ergodicity than for the 
other billiards. 
     
We have computed $S_1(E,\chi_{D_i})$ for five small domains, 
see \figref{fig:billiard-domains}c), by using the 
first 6000 eigenfunctions up to energy $E\approx 32 000$, and 
for three larger domains, see \figref{fig:cardi-qerg-S1-zusatz1}, 
by using the first 2000 eigenfunctions. 
The results are displayed in figs.~\ref{fig:cardi-qerg-S1} and 
\ref{fig:cardi-qerg-S1-zusatz1}. To determine the rate a fit of  
$S_1^{\text{fit}}(E) = \ALPHA E^{-1/4+\varepsilon}$ has been performed, 
and the resulting values for $\ALPHA$ and $\varepsilon$ are listed in 
table \ref{tab:qerg-S1-cardi}. 

We find that domain $D_3$ gives
the lowest rate of quantum ergodicity 
for the small domains $D_1$--$D_5$. 
This is caused by a considerable  number of eigenfunctions
showing an enhanced probability 
as in \figref{fig:stad-waves}e) along the vertical orbit $\overline{AB}$.
For domains $D_1$, $D_2$ we also find a slower rate than
for the other regions $D_4$, $D_5$;  in this case the 
slower rate seemingly cannot be attributed to one type  of localized 
eigenfunctions. 

\BILD{tbh}
     {
       \vspace*{1ex}

       \PSImagxy{cardi_area_kumulativ1.ps}{16cm}{11cm}\hspace*{1cm} 
     }
     {Plot of $S_1(E,{\chi_{D_i}})$ for different domains $D_i$ for 
      the cardioid billiard using the first 6000 eigenfunctions,
      see fig.~\ref{fig:billiard-domains}a) for the location 
      of the domains $D_i$.
      The inset
      shows the same curves in double--logarithmic representation 
      together with a fit of 
      $S_1^{\text{fit}}(E) = \ALPHA E^{-1/4+\varepsilon}$, 
      eq.~\eqref{eq:S1-fit-fct}.}
     {fig:cardi-qerg-S1}

The larger domains show a slower rate than the small domains, 
but the rate is not monotonically decreasing with the area of the domain. 
The rate for the largest domain $D_8$ is even of the same order of 
magnitude than the one for $D_3$, especially if one takes the 
smaller energy range for $D_8$ into account. 
This slower rate
is probably caused by the existence of different not quantum-ergodic 
subsequences with quantum limits $\qlim_k$ 
in different regions of the billiard. 
For each of the domains the influence 
of these subsequences is different 
and therefore one observes different rates.

\begin{table}[b]
  \vspace*{1cm}
  \begin{center}
   \renewcommand{\arraystretch}{1.25}
  \begin{tabular}{|c|c||c||c|}\hline
    domain & rel. area & $\varepsilon$ & $\ALPHA$    \\ \hline
    1      & 0.01722   & $ +0.047$     &  0.028      \\          
    2      & 0.01722   & $ +0.039$     &  0.037      \\          
    3      & 0.01722   & $ +0.064$     &  0.046      \\           
    4      & 0.01722   & $ +0.007$     &  0.048      \\
    5      & 0.01722   & $ +0.009$     &  0.042      \\\hline            
  6  & 0.18674   & $ +0.098$     &  0.125      \\          
  7  & 0.33104   & $ +0.115$     &  0.140      \\           
  8  & 0.50930   & $ +0.071$     &  0.213      \\   \hline         
\end{tabular}
\end{center}
  \Caption{Rate of quantum ergodicity obtained from a fit of 
           $S_1^{\text{fit}}(E) = \ALPHA E^{-1/4+\varepsilon}$
           to $S_1(E,{\chi_{D_i}})$ 
           for the cardioid billiard with domains $D_i$ as 
           shown in figs.~\ref{fig:billiard-domains}a) and 
           \ref{fig:cardi-qerg-S1-zusatz1}.
           }{tab:qerg-S1-cardi}
\end{table}

\BILD{tbh}
     {
       \vspace*{0.75cm}
        \PSImagxy{cardi_area_kumulativ1_zusatz1.ps}{16cm}{10.0cm}\hspace*{1cm}
     }
     {Plot of $S_1(E,{\chi_{D_i}})$ for larger domains
         for the cardioid billiard using the first 2000 eigenfunctions.
      Also shown are fits to eq.~\eqref{eq:S1-fit-fct}
      for the corresponding energy regions.
}
     {fig:cardi-qerg-S1-zusatz1}

A quantitative test in a similar way as for the other billiards
using a model for $S_1(E,A)$
is very difficult,  because the deviations from the
conjectured optimal rate is not only due to one subsequence. 
But the results for $D_4$ and $D_5$ clearly shows 
that here as well one has a density one subsequence of 
quantum-ergodic eigenfunctions with
rate $S_1'(E,\chi_D)\sim \slim '(D)E^{-1/4}$. 
We hope to return to 
the problem of determining the 
not quantum-ergodic subsequences and their quantum limits in the future.

The cardioid billiard is the only system we have studied to which 
the result \eqref{eq:rate-hypsyst} should be applicable. 
But for most of the domains the rate is much slower than the 
predicted one. Only the domains 4 and 5 show the expected rate. 
Therefore we have computed for these domains the factor $\classvar$ 
in eq.~\eqref{eq:rate-hypsyst}.
For the computation of $\rho(A)$ the variance of 
$\langle \chi_{D_i} \rangle_l -\frac{\Vol(D_i)}{\Vol(\Omega)}$
as a function of $l$ has been computed
using trajectory segments of length $l$
of a generic trajectory $\{q(t)\}$.
The quantity 
$ \langle \chi_{D_i} \rangle_l= \tfrac{1}{l}\int_0^l \chi_{D_i}(q(t)) \, \ud t$ 
is the relative length of the trajectory segment
lying in the domain $D_i$. By ergodicity we have
$ \lim_{l\to\infty} \langle \chi_{D_i} \rangle_l = 
\frac{\Vol(D_i)}{\Vol(\Omega)}$.
The variance of 
$\langle \chi_{D_i} \rangle_l -\frac{\Vol(D_i)}{\Vol(\Omega)}$
decreases like $\rho(A) l^{-1}$.

Using the corresponding results in equation \eqref{eq:rate-hypsyst}
we obtain
$\SIIsc(E,\chi_{D_4})=0.0062\, E^{-1/2}$ and 
$\SIIsc(E,\chi_{D_5})=0.0074\, E^{-1/2}$.
These numbers have to be compared with the result
of a fit $S_2^{\text{fit}}(E,A)$ to $S_2(E,\chi_{D_i})$. 
We obtain
$S_2^{\text{fit}}(E,\chi_{D_4})=0.0036\, E^{-0.47}$
and
$S_2^{\text{fit}}(E,\chi_{D_5})=0.0031\, E^{-0.48}$.
On sees that the theoretical prediction is too large
by a factor of approximately 2.
This deviation might be related to the factor $g$ in 
\eqref{eq:rate-hypsyst}, which counts the 
mean multiplicities in the classical length spectrum. In the cardioid 
billiard the asymptotic value $g=2$ is reached very late, for the 
shorter periods one rather has $g\approx 1$, which would 
lead to a better agreement of eq.~\eqref{eq:rate-hypsyst} with the data 
for $D_4$ and $D_5$.

For a better understanding it seems necessary to check in
detail, whether any of the assumptions leading to eq.~\eqref{eq:rate-hypsyst}
is not fulfilled for the 
domains of the cardioid billiard.
It would also be very interesting to
investigate if the slower rates can be described using the
expression in terms of the classical correlation function.
We will leave these questions for a separate study.

%%%%%%%%%%%%%%%%%%%%%%%%%%%%%%%%%%%%%%%%%%%%%%%%%%%%%%%%%%%%%%%%%%%%%%%%%%%
\subsubsection{The influence of the boundary} \label{sec:boundary-effects}
%%%%%%%%%%%%%%%%%%%%%%%%%%%%%%%%%%%%%%%%%%%%%%%%%%%%%%%%%%%%%%%%%%%%%%%%%%%

\BILD
     {tbh}
     {
       \begin{center}
       \begin{minipage}{7cm}
         \hspace*{-2.0cm}\PSImagx{stad18dd_sum250.ps}{9cm}

         \vspace*{-7cm}  a) \vspace*{7cm}
       \end{minipage}
       \hspace*{-1cm}
       \begin{minipage}{6cm}
         \PSImagxy{stad18dd_sum250_schnitt.ps}{8cm}{7cm}

         \PSImagxy{stad18dd_sum1000_schnitt.ps}{8cm}{7cm}
       \end{minipage}
       \end{center}
     }
     {In a) we show a three dimensional plot of the sum 
      $\efsum(x,y)=\frac{1}{N(E)} \sum_{E_n\le E} |\psi(x,y)|^2$ 
      involving the first $250$ eigenfunctions of the $a=1.8$ 
      stadium with odd--odd symmetry.
      The  pictures on the right show a cross section $\efsum(1,y)$ for 
      using the first b) 250 and c) 1000 eigenfunctions.
      The dashed curves in b) and c) display the evaluation using the 
      first two terms in formula \eqref{eq:Hoermander}.
      These results are used to explain the fast rate in the low energy range 
      for the stadium billiard for large domains.
     }
     {fig:wfk-sum-stadion}

In all three billiards  we observe the phenomenon, that 
for large domains 
$S_1(E,\chi_D)$ 
decays faster
at low energies than at high   
energies. This can be seen in \figref{fig:cos-qerg-S1-zusatz1} 
for domain $D_8$ in the cosine billiard, in 
\figref{fig:stad-qerg-S1-large-domain} for domains 
$D_7$ and $D_8$ in the stadium billiard and in 
\figref{fig:cardi-qerg-S1-zusatz1} for domains $D_6$, $D_7$ and $D_8$ 
in the cardioid billiard. The other large domains 
we studied showed the same behavior. The only exceptions are 
the domains $D_9$ in the cosine billiard and in the stadium billiard, 
which consist of the whole rectangular part. For these domains 
no faster rate at low energies is visible.

Qualitatively this behavior can be understood by the vanishing 
of the probability density $|\psi_n(q)|^2$ of the 
eigenstates at the boundary due to the Dirichlet boundary conditions. 
Because of the normalization of $\psi_n(q)$
the reduced probability density at the boundary has to be compensated 
by an enhancement of the probability density  inside the billiard, 
which leads to larger oscillations of the probability 
density near the boundary. 

Let us assume that this compensation of the 
probability density  takes place in a strip along 
the boundary of a few de Broglie wavelength width. Then the integral of 
the probability density $|\psi_n(q)|^2$ over 
a domain $D$  feels the influence of the boundary only up 
to a certain energy, proportional to the inverse 
square of the distance between $D$ and the boundary $\partial\Omega $. 
Furthermore the 
boundary influence will be proportional to the overlap of $D$ and 
the strip at the boundary. This overlap decreases like 
$1/\sqrt{E_n}$, and therefore $S_1(E,\chi_D)$ should decrease
with such a rate at low energies.
So the assumption that the 
compensation takes place in a small strip along the boundary 
leads exactly to the behavior we observe. Moreover 
a domain like $D_9$ which extends to the boundary $\partial\Omega $ 
will not feel any influence, because  
the boundary effect is  compensated entirely inside this domain.

To justify our assumption 
on the range of the boundary influence we refer to the following 
result  on the asymptotic  
behavior of  the  summed probability densities on  a two-dimensional 
Riemannian 
manifold with $C^{\infty}$-boundary 
\cite[Theorem 17.5.10]{Hoe85a};
\begin{equation} \label{eq:Hoermander}
  \sum_{E_n\le E} \left| \psi_n(q) \right|^2 
        = \frac{1}{4\pi} E - \frac{1}{4\pi} 
          \frac{J_1 \left(2d(q) \sqrt{E}\right)}{d(q)} 
          \sqrt{E} +  R(q,E) \;\;,
\end{equation}
where $d(q)$ is the shortest distance of the point $q\in\Omega$
to the boundary. The remainder $R(q,E )$ satisfies the estimate 
$|R(q,E )|\leq C\sqrt{E}$.
The second term in \eqref{eq:Hoermander} describes the 
influence of the boundary, for 
$d(q)\to 0$  the term  tends to $-E/(4\pi )$ and cancels the 
contribution from the first term, such that the boundary 
conditions are fulfilled. 
In \figref{fig:wfk-sum-stadion}a) the normalized sum 
\begin{equation}\label{eq:meanwave}
  \efsum(x,y) = \frac{1}{N(E)}  \sum_{E_n\leq E}|\psi_n(x,y)|^2\,\, ,
\end{equation}
is displayed
for the stadium billiard, using the first $250$ eigenfunctions. 
One nicely sees how the probability density is forced to vanish 
at the boundary, and how the compensation leads to large oscillations 
near the boundary. In \figref{fig:wfk-sum-stadion} b) and c)
we show two cross sections through the function (\ref{eq:meanwave})
at two different energies, and compared it to the result one 
gets from the first two terms on the right hand 
side of   (\ref{eq:Hoermander}). 
The agreement is quite impressive, especially 
near the boundary ($y=0$ and $y=1$). 
So although the stadium billiard does not 
have $C^{\infty}$-boundary, the result \eqref{eq:Hoermander} seems to remain 
valid. 
One furthermore observes that  with higher 
energies the $y$-range 
on which  the agreement is excellent increases.

The averaged probability density (\ref{eq:meanwave}) shows exactly the 
behavior we assumed for the individual wavefunctions, in 
order to explain the fast rate of quantum ergodicity at low energies 
for domains near the boundary. The influence of the 
Dirichlet boundary condition is 
concentrated near the boundary, and it decays at a length scale proportional 
to the de Broglie wavelength. 
So with the help of eq.~\eqref{eq:Hoermander} one gets a good 
qualitative understanding of the boundary influence on the rate of 
quantum ergodicity. 

In order to try to get a quantitative understanding we used 
eq.~\eqref{eq:Hoermander} to derive as in \cite{AurTag97:p} a 
mean eigenfunction which incorporates the boundary influence 
\begin{equation}\label{eq:mod-wave}
|\psi_{n}(q)|^2\approx  \frac{1}{\Vol(\Omega) -
\frac{\Vol(\partial\Omega)}{2\sqrt{E_n}}}\, 
\left(1-J_0(2d(q)\sqrt{E_n})\right) \,\, .
\end{equation}
Integrating this expression over a domain $D$ should give  
for the expectation values  
the mean value plus the corrections due to the boundary of 
$\chi_D$. 
By incorporating this into $S_1(E,\chi_D)$ 
one obtains an expression, which we compared
with our numerical data. 
Although \eqref{eq:mod-wave} implies a faster decay rate 
at low energies, it is not as strong as the numerically
observed one.
This deviation must be caused by considerable 
fluctuations of the boundary influence on the individual states $\psi_n$ 
around the mean influence described by 
\eqref{eq:Hoermander} and \eqref{eq:mod-wave}.

%%%%%%%%%%%%%%%%%%%%%%%%%%%%%%%%%%%%%%%%%%%%%%%%%%%%%%%%%%%%%%%%%%%%%%%%%%%%%%%
\subsection{Quantum ergodicity in momentum space}
%%%%%%%%%%%%%%%%%%%%%%%%%%%%%%%%%%%%%%%%%%%%%%%%%%%%%%%%%%%%%%%%%%%%%%%%%%%%%%%

Up to here
we have investigated the behavior of the wavefunctions in 
position space only. Now we turn our attention to the rate 
of quantum ergodicity in momentum space, which 
is studied here for the first time numerically. 

Quantum ergodicity predicts that the angular distribution of the 
momentum probability distribution $|\widehat{\psi}_n(p)|^2$ tends 
to $1/(2\pi)$
in the weak sense, see eq.~\eqref{eqn:qet-ft-version}.
Therefore we study an observable with symbol 
$\chi_{C(\theta,\Delta\theta)}(p)$ whose
expectation value gives the probability of finding the particle 
with momentum-direction in the interval 
$]\theta -\Delta\theta /2,\theta +\Delta\theta /2[$. Recall that 
$\chi_{C(\theta,\Delta\theta)}(p)$ denotes
the characteristic function of the circular sector 
$C(\theta,\Delta\theta)=\{p\in\R^2\, |\, \arctan (p_y/p_x)\in ]\theta -\Delta\theta /2,\theta +\Delta\theta /2[\}$, and the classical mean value 
of $\chi_{C(\theta,\Delta\theta)}(p)$ is  $\Delta\theta/(2\pi )$.

Only eigenfunctions of odd parity of the  not desymmetrized systems 
are considered here due to our method of 
computing the Fourier transformation directly from 
the normal derivative $u_n(\omega)$ of the eigenfunction 
$\psi_n(q)$. From Green's theorem 
one easily finds the formula   
\begin{equation}
\widehat{\psi}_n(p)=\frac{1}{p^2-E_n}\,\frac{1}{2\pi}
\Int_{\partial\Omega}\ue^{-\ui q(\omega )p}\,u_n(\omega)\, \ud\omega \,\, ,
\end{equation}
where $q(\omega )$ denotes a point on the boundary $\partial\Omega$.
The advantage of this formula is that 
it allows to compute 
the Fourier transform directly from $u_n(\omega)$, which 
can be obtained using the boundary integral method.
For desymmetrized systems, like the ones considered here, one uses 
an appropriate Greens function which vanishes at the 
lines of symmetry, and therefore removes them from the 
boundary integral, see e.g.~\cite{SieSte90b}.
 This reduces the computational effort, 
but one does not get the normal derivatives on these parts of the 
boundary of the desymmetrized system.
Therefore our results for the rate of quantum ergodicity 
in momentum space  are
sufficient to rule out the 
possibility of a totally different behavior in momentum 
space than in position space. Since the rate for all 
eigenfunctions cannot be faster than the one for a subsequence of 
positive density, we get a lower bound for the 
rate of the full system.

The time reversal invariance leads for the Fourier transformed 
eigenfunctions to the symmetry 
$\widehat{\psi}_n(-p)=\overline{\widehat{\psi}}_n(p)$. 
Therefore $|\widehat{\psi}_n(-p)|^2=|\widehat{\psi}_n(p)|^2$,  
and this reduces the angle 
interval we have to study to 
$[0,\pi[$. The additional reflection symmetries  in the considered 
billiards  further reduce the 
relevant angle interval  to $[0,\pi /2[$.

For our numerical computations we have chosen five equidistant intervals,
centered at $\theta_i=(i-1/2) \frac{\pi}{10}$ with $i=1,\ldots,5$
of width $\Delta \theta=\frac{\pi}{10}$.

As in the case of quantum ergodicity in coordinate space, see 
eq.~\eqref{eq:an-diff} and fig.~\ref{fig:stad-qerg-area}, 
one observes large fluctuations of
$\langle \psi_n ,\chi_{C(\theta,\Delta\theta)}
\psi_n\rangle-\Delta\theta/(2\pi ) $
around 0.
Therefore we again consider the cumulative
version \eqref{eqn:qet-sum-version}
of the quantum ergodicity theorem, which reads in this case 
\begin{equation} 
  S_1(E,\Op{\chi_{C(\theta,\Delta\theta)}}) = \frac{1}{N(E)} \sum_{E_n\le E} 
            \Bigg| \; \Int_{C(\theta,\Delta\theta)} 
                    |\widehat{\psi}_{n}(p)|^2 \; \ud p
                    - \frac{\Delta\theta}{2\pi}\Bigg| 
     \to 0 \quad \text{ for } E\to \infty\;\; .
\end{equation}

\begin{figure}[tbh]
     {
%       \vspace*{-1.25cm}

      % \hspace*{0.5cm}\PSImagxy{stad_four_kumulativ.ps}{16cm}{10.0cm}
     }

     \Caption{Plot of $S_1(E,\Op{\chi_{C(\theta_i,\Delta\theta)}})$ 
              for $\theta_i=(i-1/2) \frac{\pi}{10}$ with $i=1,\ldots,5$
              and $\Delta \theta=\frac{\pi}{10}$
              for the stadium  billiard using the first
              2000 eigenfunctions.}
     {fig:ft_vert_serie-stadium}
     \vspace*{1ex}

     {
       % \hspace*{0.5cm}\PSImagxy{cardi_four_kumulativ.ps}{16cm}{10.0cm}
     }

     \Caption{Plot of $S_1(E,\Op{\chi_{C(\theta_i,\Delta\theta)}})$ 
              for $\theta_i=(i-1/2) \frac{\pi}{10}$ with $i=1,\ldots,5$
              and $\Delta \theta=\frac{\pi}{10}$
              for the cardioid billiard using the first
              2000 eigenfunctions.}
     {fig:ft_vert_serie-cardioid}
\end{figure}

The results for $S_1(E,\Op{\chi_{C(\theta_i,\Delta\theta)}})$
are shown in \figref{fig:ft_vert_serie-stadium}
for the stadium billiard and
in \figref{fig:ft_vert_serie-cardioid}
for the cardioid billiard. 
In each case 2000 eigenfunctions have been used.
For the cardioid billiard the inset shows 
a double logarithmic representation
together with the fits of $S_1^{\text{fit}} (E)$, eq.~\eqref{eq:S1-fit-fct}.
For the cosine billiard no computations
of the rate $S_1(E,\Op{\chi_{C(\theta_i,\Delta\theta)}})$
in momentum space have been performed.

\begin{table}[tb]
   \renewcommand{\arraystretch}{1.25}
  \begin{center}
  \begin{tabular}{|c|c|c|}\hline
system       & domain & $\varepsilon$ \\\hline

             & 1    & $0.15 $   \\ 
             & 2    & $0.12 $   \\ 
stadium      & 3    & $0.15 $   \\ 
             & 4    & $0.09 $   \\
             & 5    & $0.18 $   \\ \hline 
%%%%%%%%%%%%%%%%%%%%%%%%%%%%%%%%%%%%%%%%%%%%%%%%%%%%%%%%%
             & 1    & $0.050 $   \\  
             & 2    & $0.075 $   \\  
cardioid     & 3    & $0.026 $   \\  
             & 4    & $0.079 $   \\  
             & 5    & $0.076 $   \\ \hline  
\end{tabular}
\end{center}
  \Caption{Rate of quantum ergodicity obtained from a fit of 
           $S_1^{\text{fit}}(E) = \ALPHA E^{-1/4+\varepsilon}$ to the 
           numerically obtained function 
           $S_1(E,\Op{\chi_{C(\theta_i,\Delta\theta)}})$
           for the different systems and angle sectors  
           $C(\theta_i,\Delta\theta)$.
           }{tab:qerg-ft-S1}
\end{table}

As in position space one expects that the rate is 
strongly influenced by not quantum-ergodic subsequences of eigenfunctions. 
For the bouncing ball modes in the stadium billiard  one has 
\begin{equation}
\lim_{E_{n''}\to\infty}\Int_{C(\theta,\Delta\theta)} 
                    |\widehat{\psi}_{n''}(p)|^2 \; \ud p =
           \begin{cases} 0\,\, &\text{for}\quad \frac{\pi}{4}\notin 
                                 \;]\theta-\Delta\theta,\theta+\Delta\theta[\\
                         1\,\, &\text{for}\quad \frac{\pi}{4}\in 
                                 \;]\theta-\Delta\theta,\theta+\Delta\theta[
           \end{cases}\,\, ,
\end{equation}
and so the coefficient $\slimbb ''(\chi_{C(\theta_i,\Delta\theta)})$ 
in the model \eqref{mod-rate} for $S_1(E,\chi_{C(\theta_i,\Delta\theta)})$
is given by 
\begin{equation}
\slimbb ''(\chi_{C(\theta_i,\Delta\theta)})=
   \begin{cases} \frac{1}{20}\,\,  & \text{for}\quad i=1,\ldots,4 \\
                 \frac{19}{20}\,\, & \text{for}\quad i=5
   \end{cases}\,\, .
\end{equation}

The results for the rate of quantum ergodicity, characterized
by $\varepsilon$, are listed in table \ref{tab:qerg-ft-S1}.
It turns out that the rate is  slower than
the rate of quantum ergodicity for the small domains in configuration space.
Moreover the agreement of $S_1(E,\Op{\chi_{C(\theta_i,\Delta\theta)}})$ 
with the fit is not as good as in the case of $S_1(E,\chi_D)$,
in particular the fluctuations 
of $S_1(E,\Op{\chi_{C(\theta_i,\Delta\theta)}})$ 
are much larger than in position space.

In the stadium billiard the interval $5$, which corresponds
the direction of the bouncing ball orbits,  shows the slowest rate. 
But as we already noted in the discussion of the rate in position space, 
the bouncing ball modes alone cannot cause such a slow rate, because their 
counting function increases only as $E^{3/4}$.
So a considerable number of the additional 
not quantum-ergodic states which are responsible for the slow rate in 
position space must also have an enhanced momentum density around $\pi/2$.  
But the slow rates for the other angular intervals
indicate that not all 
not quantum-ergodic states show this behavior in momentum space.   

For both billiards one observes that the order of magnitude of 
$\varepsilon$ in momentum space is the same as for the large domains in 
position space. 
Therefore the results are compatible with the results in  position space,
but the large fluctuations indicate that one has to go higher in the 
energy in momentum space than in position space.

%%%%%%%%%%%%%%%%%%%%%%%%%%%%%%%%%%%%%%%%%%%%%%%%%%%%%%%%%%%%%%%%%%%%%%%%
\subsection{Fluctuations of expectation values}\label{sec:fluct-of-exp-values} 
%%%%%%%%%%%%%%%%%%%%%%%%%%%%%%%%%%%%%%%%%%%%%%%%%%%%%%%%%%%%%%%%%%%%%%%%

Another aspect of great interest
is how the expectation values 
$\langle \psi_n,\OpA \psi_n\rangle$
fluctuate around their mean value $\overline{\sigma(\OpA)}$.
Since the mean fluctuations decrease for large $n$,
one has to consider the distribution of
\begin{equation}\label{eq:norm-fluc}
  \xi_n=\frac{\langle \psi_n,\OpA \psi_n\rangle-\overline{\sigma(\OpA)}}
           {\sqrt{\tilde{S}_2(E_n,\OpA)}} \;\;.
\end{equation}
Here $\tilde{S}_2(E,\OpA)=\Xi \, S_2(E,\OpA)$
with $\Xi$ being a correction 
necessary to ensure that the distribution of $\xi_n$ has
unit variance; see below for an explanation.
So the question is whether a limit distribution $P(\xi)$
of $\xi_n$ exists in the weak sense, i.e.\
\begin{equation}\label{def:limit-distribution}
  \lim_{N\to\infty} \frac{1}{N} \sum_{n=1}^N g(\xi_n) = 
      \Int_{-\infty}^\infty g(\xi) P(\xi) \; \ud \xi \;\;,
\end{equation}
where $g(\xi)$ is a  bounded  continuous function.
It is natural to conjecture that this distribution tends
to a Gaussian normal distribution,
\begin{equation}
  \label{eq:normal-distribution}
  P(\xi) = \frac{1}{\sqrt{2\pi}} \exp(-\xi^2/2) \;\;,
\end{equation}
as in random matrix theory \cite[section VII]{BroFloFreMelPanWon81}. 
Note that this is a conjecture for every observable, i.e.\ 
the asymptotic distribution should be independent of
the special observable under investigation.
For hyperbolic surfaces a  study of $P(\xi)$ for an observable in 
position space is contained in \cite{AurTag97:p}, where a good 
agreement with a Gaussian normal distribution was observed. 
In \cite{EckFisKeaAgaMaiMue95} $P(\xi)$ was studied for the Baker's map 
and the hydrogen atom in a strong magnetic field, and a fair agreement with 
a Gaussian was found.

\BILD{htb}
     { 
      \begin{center}
         \vspace*{-1.0cm}
         \PSImagxy{stad_gauss_4_cum.ps}{14cm}{9.75cm}

         \vspace*{2.5ex}
 
         \PSImagxy{odd_n__gauss_4_cum_special.ps}{14cm}{9.75cm}
      \end{center} 

     }
     {Cumulative distribution of 
      $\xi_n=(\langle \psi_n,A\psi_n\rangle -
      \overline{\sigma (A)})/\sqrt{\tilde{S}_2(E_n,A)}$ for the stadium billiard
      for domain 4, $A=\chi_{D_4}$, and for the cardioid billiard 
      with observable $A=\chi_{D_4} - \chi_{D_5}$. 
      In both cases we haven chosen $n\in[2000,6000]$.
      The dashed curve
      corresponds to the cumulative normal distribution.
      The insets show the distribution of $\xi_n$ together with the normal
      distribution with zero mean and unit variance,
      eq.~\eqref{eq:normal-distribution} (dashed curve).}
     {fig:qerg-gauss}

\BILD{!t}
     { 
     \vspace*{1ex}
      \begin{center}
         \PSImagxy{odd_n__ft_vert_serie_5_gauss_3_cum.ps}{14cm}{10cm}
      \end{center} 

     }
     {Cumulative distribution of 
      $\xi_n=(\langle \psi_n,A\psi_n\rangle -
      \overline{\sigma (A)})/\sqrt{\tilde{S}_2(E_n,A)}$ for the cardioid billiard,
      for the observable $\chi_{C(\theta,\Delta\theta)}(p)$ 
      in momentum space with $\theta=5\pi/20$ and $\Delta \theta=\pi/10$.
      The dashed curve
      corresponds to the cumulative normal distribution.
      The insets show the distribution of $\xi_n$ together with the normal
      distribution with zero mean and unit variance,
      eq.~\eqref{eq:normal-distribution} (dashed curve).}
     {fig:ft-qerg-gauss}

However, already from the plots of $d_i(n)$ shown in 
\figref{fig:stad-qerg-area}
it is clear that the fluctuations
are not symmetrically distributed around zero, but
have more peaks with large positive values.
The reason is  that $d_i(n)= \langle \psi_n,\chi_{D_i} \psi_n\rangle-
\tfrac{\Vol(D_i)}{\Vol(\Omega )}$ 
has to satisfy the inequality
\begin{equation}
  -\frac{\Vol(D_i)}{\Vol(\Omega )} 
    \le \langle \psi_n,\chi_{D_i} \psi_n\rangle-\frac{\Vol(D_i)}{\Vol(\Omega )} 
    \le 1 -\frac{\Vol(D_i)}{\Vol(\Omega )} \;\;.
\end{equation}
This already indicates 
that the approach to an asymptotic Gaussian behavior could be 
rather slow. Therefore we have tested additionally  
for the cardioid billiard the observable 
$A=\chi_{D_4}-\chi_{D_5}$ where the expectation values 
fluctuate symmetrically around zero, and one expects a faster approach 
to a Gaussian behavior.  
In \figref{fig:qerg-gauss}a) we show
the  cumulative distribution  
\begin{equation}
  I_N(\xi) = \frac{1}{N}\;  \#\left\{ n \setsep \xi_n < \xi \right\}
\end{equation}
for domain $D_4$ of the stadium billiard
and in \figref{fig:qerg-gauss}b)        
$I_N(\xi )$ is shown for 
the observable $A=\chi_{D_4}-\chi_{D_5}$ in case of
the cardioid billiard.
In both cases all values of $\xi_n$ with $n\in [2000,6000]$
have been taken into account, giving $N=4000$.
For the rate $S_2(E,\chi_D)$ we used the result of a fit
to $S_2^{\text{fit}}(E) = \ALPHA E^\alpha$.
The insets show the corresponding distributions of $\xi_n$
in comparison with the normal distribution, eq.~\eqref{eq:normal-distribution}.
Notice, that no further fit of the mean or the variance of
the Gaussian has been made.
Example a) is the case for which we have found 
the worst agreement with a Gaussian
(of all the small domains we have tested).
The observable chosen for b) 
gives very good 
agreement with the Gaussian distribution.
In the case of $\chi_{D_4}$ in the stadium billiard there is a 
significant peak
around $\xi=-2$, which is due to the bouncing ball modes, for
which $\langle \psi_{n''}, \chi_{D_4} \psi_{n''} \rangle$
is approximately zero, see \figref{fig:stad-qerg-area}. 
Therefore one has a larger fraction
with negative $\xi_{n''}$.
For the distribution in case of the observable
$A=\chi_{D_4}-\chi_{D_5}$  of the cardioid billiard
we obtain a significance niveau of $23\%$ for the Kolmogorov--Smirnov test
(see e.g.\ \cite{PreTeuVetFlan92})
with respect to the cumulative normal distribution.

We also studied the distribution of $\xi_n$  
for the observables 
$a(p,q)=a(p)=\chi_{C(\theta,\Delta\theta)}(p)$
in momentum space.
For the stadium billiard the computed distributions show in the considered
energy range clear deviations from a Gaussian,
as one already expects from fig.~\ref{fig:ft_vert_serie-stadium}.
The best result was obtained for the cardioid billiard
for the interval given by $i=3$ 
(with $\theta_i= (i-1/2) \frac{\pi}{10}$ and $\Delta \theta=\frac{\pi}{10}$)
and is shown in \figref{fig:ft-qerg-gauss}.
The agreement is quite good, the Kolmogorov--Smirnow test
gives a significance niveau of $29\%$.

There is one subtle point concerning the variance of the distribution of $\xi_n$.
Since $S_2(E,A)$ does not represent a local variance around $E$, but
a global one, it is necessary to take this into account to obtain
for the fluctuations
a variance of unity. If the rate behaves as $S_2(E,A)=a E^\alpha$
then the correction $\Xi$ is given by $\Xi=\alpha+1$,
e.g.\ for $\alpha=-1/2$ we have $\Xi=1/2$.
See \cite{AurBaeSte97} for a more detailed discussion
on this point in the case of the distribution of
the normalized mode fluctuations.

Let us now discuss the influence of not quantum-ergodic
sequences to the possible limit distribution.
Assume that the rate for the quantum-ergodic states is 
$S_2 '(E,A)\sim a E^{-\alpha}$ with some
power $\alpha$. If we have a subsequence of
not quantum-ergodic states such that the total rate
is 
$S_2(E,A)\sim a' E^{-\alpha'}$,
we can have two different situations,
either  $\alpha=\alpha'$, or
$\alpha<\alpha'$. In the first case the 
not quantum-ergodic states have no influence
on the limit distribution. 
In the second case where the not quantum-ergodic states dominate the rate,
the normalization by a rate which is slower  than
the one of the quantum-ergodic subsequence implies
that  we have $P(\xi)=\delta(\xi)$. 

If one instead normalizes  the fluctuations with the rate of the 
quantum-ergodic subsequence, $S_2'(E,A)$, 
\begin{equation}
\tilde{\xi}_n:=\frac{\langle \psi_n,\OpA \psi_n\rangle-\overline{\sigma(\OpA)}}
           {\sqrt{\tilde{S}_2'(E_n,\OpA)}}\,\, ,
\end{equation}
with $\tilde{S}_2'(E_n,\OpA) =\Xi S_2'(E,A)$, then the limit distribution 
is determined only by the quantum-ergodic subsequence,
independent of the behavior of the not 
quantum-ergodic subsequence. To see this we split 
\eqref{def:limit-distribution} into the different parts 
\begin{equation}
\frac{1}{N(E)}\sum_{E_n\leq E}g(\tilde{\xi}_n)=
\frac{N'(E)}{N(E)}\frac{1}{N'(E)}\sum_{E_{n'}\leq E}g(\tilde{\xi}_{n'})+
\frac{N''(E)}{N(E)}\frac{1}{N''(E)}\sum_{E_{n''}\leq E}g(\tilde{\xi}_{n''})\,\, .
\end{equation}
Since $\lim_{E\to\infty}\frac{N'(E)}{N(E)}=1$, 
$\lim_{E\to\infty}\frac{N''(E)}{N(E)}=0$ and 
$\frac{1}{N''(E)}\sum_{E_{n''}\leq E}g(\tilde{\xi}_{n''})\leq \max_{\xi\in\R}g(\xi )$, 
one gets 
\begin{equation}
\lim_{E\to\infty}\frac{1}{N(E)}\sum_{E_n\leq E}g(\tilde{\xi}_n)=
\lim_{E\to\infty}\frac{1}{N'(E)}\sum_{E_{n'}\leq E}g(\tilde{\xi}_{n'})\,\, .
\end{equation}
We conjecture that the fluctuations of the quantum-ergodic subsequence is 
Gaussian, 
and therefore all fluctuations, when normalized with the rate of the 
quantum-ergodic subsequence, are Gaussian.

%%%%%%%%%%%%%%%%%%%%%%%%%%%%%%%%%%%%%%%%%%%%%%%%%%%%%%%%%%%%%%%%%%%%%%%%%%%%%%%

\section{Summary}

The aim of the present paper is to give
a detailed study of the rate of 
quantum ergodicity in Euclidean billiards. 
We first have given a short introduction to the 
quantum ergodicity theorems in terms of pseudodifferential
operators. 
We have shown that the quantum ergodicity theorems
of Shnirelman, Zelditch, Colin de Verdi\'ere and others
are equivalent to a weak from of the semiclassical
eigenfunction hypothesis for ergodic systems put
forward in \cite{Vor76,Vor77,Ber77b,Ber83}.
That is, the quantum ergodicity theorem is equivalent to the statement
that for ergodic systems the Wigner functions $W_n (p,q)$ fulfill
\begin{equation}
\W_{n_j} (p,q )\sim \frac{1}{\text{Vol}(\Sigma_{E_{n_j}} )} \, \delta (H(p,q)-E_{n_j}) \,\, ,
\end{equation}
for $E_{n_j}\to\infty$ and $\{n_j\}\subset\N$ a subsequence of 
density one. 

Of great importance for the practical applicability
of the quantum ergodicity theorem is the question,
at which rate quantum mechanical expectation values
$\langle \psi_n , A \psi_n \rangle$ tend to their mean value 
$\overline{\sigma(A)}$.
Different arguments were presented previously in favor of an expected 
rate of quantum ergodicity $S_1(E,\OpA) = O(E^{-1/4+\varepsilon})$,
for all $\varepsilon>0$, in the case of strongly chaotic systems.
In section \ref{sec:rate-of-quantum-ergodicity}
we discussed the influence of not quantum-ergodic
subsequences to the rate. If their counting function
increases sufficiently fast, they
can dominate the behavior of 
$S_1(E,\OpA)$ asymptotically. Together with results 
from \cite{BaeSchSti97a} for the number of bouncing ball modes 
in certain billiards it follows that one can find for arbitrary 
$\delta>0$ an ergodic billiard for which $S_1(E,\OpA)=O(E^{-\delta})$. 
That is, the quantum ergodicity theorem gives a sharp bound,
which cannot be improved without additional assumptions on the system.

We furthermore developed a simple but powerful model for the behavior of 
$S_1(E,\OpA)$ in the presence of not quantum-ergodic eigenfunctions, 
whose main ingredient is that the quantum-ergodic eigenfunctions 
should obey the optimal rate $E^{-1/4}$. 
The discussion shows that the total rate of quantum ergodicity
can be strongly influenced by those exceptional subsequences.
Not only that they can cause the  rate  to be much slower than $E^{-1/4}$, 
they can lead as well to a grossly different behavior of $S_1(E,A)$ 
at low and intermediate and at high energies. 

The numerical investigations are carried out for three types
of Euclidean billiards, the stadium billiard (with different parameters), 
the cosine and the cardioid billiard.
The results are based on 2000 eigenfunctions for the cosine billiard, and 
up to  6000 eigenfunctions for the stadium and the cardioid billiard.
As observables we have used characteristic functions of different
domains in position space and also a class of 
observables in momentum space, considered here for the first time.

It turns out that the rate of quantum ergodicity in position
space is in good agreement with a power law decay,
$S_1(E,\OpA) \sim E^{-1/4+\varepsilon}$.
The difference $\varepsilon$ between the exponent and $1/4$ 
is found to be small for several domains and systems.
However we also find a number of significant examples showing a slow rate 
(i.e.\ $\varepsilon>0$ and not small). These are discussed
in detail and can be attributed to subsequences of localized eigenfunctions.

For the cosine billiard we find that the rate agrees well with the expected
rate, in particular for 
the small domains. 
However asymptotically the rate has to obey $S_1(E,A) \sim E^{-1/10}$,
because the counting function of the bouncing ball modes
increases as $E^{9/10}$. The asymptotic regime for the rate
lies far beyond any presently computable number of energy levels.
By incorporating the knowledge on the counting function
obtained from our previous work, we 
tested
our 
model \eqref{mod-rate} for the rate successfully
for all the considered domains.

For the stadium billiard the situation is more complicated:
here the counting function of the bouncing ball modes
increases as $E^{3/4}$ and therefore, as discussed
in sec.~\ref{sec:rate-of-quantum-ergodicity},
cannot influence the rate.
However, we  find for the stadium billiard
that the rate 
is for several domains in position space much slower than expected.
After discussing and testing several possibilities 
our explanation for this observation is
that in the stadium billiard 
there exist a much larger 
 subsequence of eigenfunctions which
have an enhanced probability density
in the rectangular part of the billiard than just the 
bouncing ball modes. 
They nevertheless have density zero, but their counting function 
increases stronger than $E^{3/4}$. 
Of course, we cannot decide whether this subsequence
either has a quantum limit different from the Liouville
measure, or if it is a quantum-ergodic subsequence,
with an exceptionally slow rate.

For the cardioid billiard we also have domains 
for which the rate is proportional to $E^{-1/4}$. But 
we also find
significant exceptions, in particular for domain $D_3$
the rate is much slower, and this can be attributed
to a number of eigenstates which show localization
along the unstable periodic orbit $\overline{AB}$.
For the cardioid billiard we also tested the result
from \cite{EckFisKeaAgaMaiMue95}, eq.~\eqref{eq:rate-hypsyst}, 
for the domains $D_4$ and $D_5$,
for which the rate is closest to the optimal rate.
However the semiclassical result does not agree 
with our numerical results for the rate.
It would be interesting to study this in more detail.

From our numerical results we obtain the following general picture:
In the studied systems there is a quantum-ergodic subsequence of 
density one whose rate is $S'_1(E,\OpA) = O(E^{-1/4+\varepsilon})$.  
If one observes a slower rate of $S_1(E,\OpA)$ by using all eigenfunctions,
this is caused by
a subsequence of density zero,
whose counting function increases stronger than $E^{3/4}$.
These exceptional eigenfunctions show localization effects and 
and probably they 
tend to some non ergodic quantum limit. However we cannot rule out the 
possibility that they are quantum-ergodic but with a much slower 
rate than the majority of eigenfunctions.

We have found
furthermore an effect due to the 
boundary conditions.
 For domains lying next to the boundary we 
observed that
the rate may be considerably faster at low energies.
The qualitative explanation of the phenomenon is
that the probability density of the eigenstates show
enhanced fluctuations near the boundary
because of the boundary conditions.

Using an observable depending only on the momentum,  we 
studied  quantum ergodicity in momentum space too, which is done 
 here for the first time to our knowledge.
We find that in general the rate of quantum ergodicity is
of the same magnitude as for the large 
domains in position space.
Furthermore the oscillations of $S_1(E,A)$ are larger in momentum space, 
which might indicate that one has to go higher in the energy in 
momentum space than in position space.

We also studied the distribution of the 
suitably normalized fluctuations 
of $\langle \psi_n,A\psi_n\rangle -\overline{\sigma(A)}$, 
see eq.~\eqref{eq:norm-fluc},
both for operators in position space and in momentum space.
For the observable $A=\chi_{D_4}-\chi_{D_5}$
in the case of the cardioid billiard we find very good agreement
with a Gaussian normal distribution
and in the case of observables depending only on the momentum
good agreement is found.
However for the stadium billiard
(and also domain $D_3$ for the cardioid billiard)
we clearly find that again subsequences of 
not quantum-ergodic states may 
have a considerable influence. 
If they dominate  $S_2(E,A)$, the distribution will tend 
to a delta function due to the normalization by $\sqrt{S_2(E,A)}$. 
But when normalizing instead 
with the rate of the quantum-ergodic states,
$\tilde{S}_2' (E,\OpA)$, we expect a universal Gaussian 
distribution of the fluctuations.

As possible investigations for the future it 
seems very interesting to study
whether the expression given in  \cite{EckFisKeaAgaMaiMue95} 
for the rate in terms of the classical correlation function
can describe our numerical results.
In particular for the cardioid billiard 
a more detailed investigation along these lines seems
promising 
as this system is the most ``generic'' one
from the three studied systems
and we find both the optimal rate and
also clear deviations.
The present paper also shows that a detailed
understanding of the phenomenon of scarred eigenfunctions
is necessary because these clearly affect the rate of quantum ergodicity.

%%%%%%%%%%%%%%%%%%%%%%%%%%%%%%%%%%%%%%%%%%%%%%%%%%%%%%%%%%%%%%%%%%%%%%%%%%%%%%%

\vspace{2ex}

{\bf Acknowledgments}

\vspace{1ex}

We would like to thank 
Dr.\ R.\ Aurich, Dr.\ J.\ Bolte, T. Hesse, Dr.\ M.\ Sieber and
Prof.\ Dr.\ F. Steiner,
for useful discussions and comments.  
Furthermore we are grateful to Prof.\ Dr.\ M. Robnik and Dr.\ T. Prosen
for the kind provision
of the eigenvalues of the cardioid billiard.  
Fig.\ref{fig:wfk-sum-stadion}a) has been visualized using
{\tt Geomview} from {\it The Geometry Center} of the University of
Minnesota and then rendered using {\tt Blue Moon Rendering Tools} 
written by L.I.~Gritz.
A.B.\ acknowledges support by the 
Deutsche Forschungsgemeinschaft under contract No. DFG-Ste 241/7-2.

\vspace{1ex}

\begin{appendix}

\section*{Appendix}

\section{Kohn-Nirenberg quantization}\label{app:relation-to-Kohn-Nirenberg}

In the mathematical literature one often prefers a 
different quantization procedure, sometimes 
called the Kohn-Nirenberg quantization \cite{Hoe85a,Fol89}, and 
the literature on quantum ergodicity often refers to this 
quantization procedure. To the 
symbol $a\in S^m(\R^2\times\Omega)$ one associates the operator 
\begin{equation}
\text{Op}^{\text{KN}} [a]f(q):=\frac{1}{(2\pi)^2}
\Int_{\R^2} \ue^{\ui pq}a(p,q)\hat{f}(p)\, \ud p\,\, ,
\end{equation}
where $\hat{f}(p):=\int_{\Omega}\ue^{-\ui qp}f(q)\, \ud q$ is 
the Fourier transform of $f$. 
The principal symbol is defined in the same way as before, i.e., 
if $a\sim\sum_{k=0}^{\infty}a_{m-k}$, then the leading term $a_m$ is called 
the principal symbol, $\sigma^{\text{KN}}(\text{Op}^{\text{KN}}[a])=a_m$. 

The usual 
quantum ergodicity theorem is now the same theorem as we have stated it, 
but with the Kohn-Nirenberg principal symbol $\sigma^{\text{KN}}$ instead 
of the principal symbol corresponding to the Weyl symbol which 
we have used. But it is well 
known, see \cite{Fol89, Hoe85a}, that if $a\in S^m(\R^n\times\R^n)$, 
then the Weyl symbol of the Kohn-Nirenberg operator belongs to the 
same symbol space, $\W [\text{Op}^{\text{KN}} [a]]\in S^m(\R^n\times\R^n)$, 
and that 
the principal symbol coincides with the Kohn-Nirenberg principal symbol, 
\begin{equation}
\sigma (\text{Op}^{\text{KN}} [a])=
\sigma^{\text{KN}}(\text{Op}^{\text{KN}}[a])\,\,  .
\end{equation}
Therefore the two formulations of the quantum ergodicity theorem 
are equivalent. 

%%%%%%%%%%%%%%%%%%%%%%%%%%%%%%%%%%%%%%%%%%%%%%%%%%%%%%%%%%%%%%
\section{Generalizations of the quantum ergodicity theorem}
\label{app:generalizations-of-the-qet}

%%%%%%%%%%%%%%%%%%%%%%%%%%%%%%%%%%%%%%%%%%%%%%%%%%%%%%%%%%%%%%

Assume we have given a quantum limit $\qlim_k$ on $\Ee$, that is 
we have a subsequence of eigenfunctions $\{\psi_{n_j}\}_{j\in\N}$, such 
that 
\begin{equation}\label{def:qlim}
\lim_{j\to\infty}\langle \psi_{n_j}, A\psi_{n_j}\rangle 
=\Int_{\Ee} \qlim_k (p,q)\sigma (A)(p,q)\, \ud\mu \,\, , 
\end{equation}
for all $A\in S^0_{\cl}(\Omega )$. 
We want to discuss the lift of $\qlim_k$ from $\Ee$ to the whole 
phase space.

To this end we express the expectation values 
for an operator of arbitrary order $m\in\R$ by the expectation values
of an operator of order zero.
This can be achieved by using  
the fact that for every $m\in\R$,  
$(-\Delta )^{\frac{m}{2}}$ is a pseudodifferential 
operator of order $m$ with principal symbol 
$\sigma \left((-\Delta )^{\frac{m}{2}}\right) = 
\left(\sigma (-\Delta )\right)^{\frac{m}{2}}=H(p,q)^{\frac{m}{2}}$, see 
\cite{ See67, See69, Tay81}. 
By multiplying an operator  $A\in S^m(\R^m)$ of order $m$ with 
the operator $(-\Delta )^{-\frac{m}{2}}$, which is of order $-m$, 
we get an operator 
$(-\Delta )^{-\frac{m}{2}}A\in S^0(\R^n)$ of order zero. For the 
expectation values of $A$ we therefore have 
\begin{equation}\label{expval:ordm}
\langle \psi_{n_j}, A\psi_{n_j}\rangle = 
E_{n_j}^{\frac{m}{2}}\langle \psi_{n_j},(-\Delta )^{-\frac{m}{2}}A \psi_{n_j}\rangle \,\, ,
\end{equation}
and on the right hand side we have an operator of order zero. 

The principal symbol of $(-\Delta )^{-\frac{m}{2}}A$ is 
according to eq.~(\ref{prod}) given by 
$\sigma ((-\Delta )^{-\frac{m}{2}})\sigma (A)=
H(p,q)^{-\frac{m}{2}}\sigma (A)$, and since by definition 
$H(p,q)=1$ on $\Ee$ we obtain from (\ref{def:qlim}) and (\ref{expval:ordm})
\begin{equation}\label{qlim:mord}
\lim_{j\to\infty}E_{n_j}^{-\frac{m}{2}}
\langle \psi_{n_j}, A\psi_{n_j}\rangle =
\Int_{\Ee} \qlim_k (p,q)\,\sigma (A)(p,q)\; \ud\mu \,\, .
\end{equation}

Thus eq.~(\ref{qlim:mord}) provides the extension
of the quantum ergodicity theorem to pseudo--differential operators
of arbitrary order $m$.

%%%%%%%%%%%%%%%%%%%%%%%%%%%%%%%%%%%%%%%%%%%%%%%%%%%%%%%%%%%%%%%%%%%%%
\section{Connection to the semiclassical eigenfunction hypothesis}
\label{app:connection-to-sc-eigenfunction-hypothesis}

%%%%%%%%%%%%%%%%%%%%%%%%%%%%%%%%%%%%%%%%%%%%%%%%%%%%%%%%%%%%%%%%%%%%%

By introducing the definition 
of the Liouville measure $\mu$, equation (\ref{qlim:mord}) can 
be written as 
\begin{equation}
\langle \psi_{n_j}, A\psi_{n_j}\rangle 
\sim E_{n_j}^{\frac{m}{2}}
\IInt \qlim_k (p,q)\sigma (A)(p,q)\frac{\delta (H(p,q)-1)}{\Vol (\Ee )}\;
\ud p\, \ud q \,\, .
\end{equation}
If one uses the homogeneity of $\sigma (A)$, i.e.\
$E_{n_j}^{\frac{m}{2}}\sigma (A)(p,q)=\sigma (A)(E_{n_j}^{\frac{1}{2}}p,q)$, 
and performs a change of the momentum coordinates from 
$p$ to $E_{n_j}^{-\frac{1}{2}}p$ one obtains 
\begin{equation} \label{eq:blubb}
\begin{split}
\langle \psi_{n_j}, A\psi_{n_j}\rangle 
&\sim
\IInt \qlim_k (E_{n_j}^{-\frac{1}{2}}p,q)\sigma (A)(p,q) \,
\frac{\delta (H(E_{n_j}^{-\frac{1}{2}}p,q)-1)}{\Vol (\Ee )}E_{n_j}^{-\frac{n}{2}}\;
\ud p\, \ud q \\
&=\IInt \qlim_k (E_{n_j}^{-\frac{1}{2}}p,q)\sigma (A)(p,q) \,
\frac{\delta (H(p,q)-E_{n_j})}{\Vol (\Ee )E_{n_j}^{\frac{n}{2}-1}}
\;\ud p\, \ud q\,\, , \\
\end{split}
\end{equation}
where furthermore the homogeneity properties of $H(p,q)$ and of the 
delta function have been used. 
In terms of the Wigner functions $W_{n_j}$ corresponding to 
$\psi_{n_j}$ 
eq.~(\ref{eq:blubb}) reads
\begin{equation}
\IInt\sigma (A)(p,q) \,W_{n_j}(p,q)\, \ud p\, \ud q \sim 
\IInt\sigma (A)(p,q)\,\qlim_k (E_{n_j}^{-\frac{1}{2}}p,q)
\,\frac{\delta (H(p,q)-E_{n_j})}{\Vol (\Ee )E_{n_j}^{\frac{n}{2}-1}}
\;\ud p\, \ud q\,\, ,
\end{equation}
where $\sigma (A)(p,q)$ can be any function
homogeneous in $p$ of degree $m$, for some arbitrary $m\in \R$.  
But since the set of all polynomials in $p$ is already dense in 
$C^{\infty}(\R^2\times \Omega)$ the set of homogeneous functions 
in $p$ is dense in  $C^{\infty}(\R^2\times \Omega)$, too. 
Therefore one gets 
\begin{equation}
W_{n_j}(p,q)\sim \qlim_k (E_{n_j}^{-\frac{1}{2}}p,q)\,
\frac{\delta (H(p,q)-E_{n_j})}{\Vol (\Ee )E_{n_j}^{\frac{n}{2}-1}}\,\, .
\end{equation}
Note that $\Vol (\Ee )E_{n_j}^{\frac{n}{2}-1}=\Vol (\Sigma_{E_{n_j}})$, 
and if we extend $\qlim_k(p,q)$ from $\Ee$ to the whole phase space by 
requiring it to be homogeneous of degree zero, 
$\qlim_k(p,q)=\qlim_k(p/\sqrt{H(p,q)},q)$, then we finally can write 
\begin{equation}
W_{n_j}(p,q)\sim \qlim_k(p,q) \frac{\delta (H(p,q)-E_{n_j})}{\Vol (\Sigma_{E_{n_j}})}
\,\, \quad \text{for}\,\,j\to\infty \,\, .
\end{equation}
for a subsequence $\{n_j\}\subset\N$ of density 1. 
This shows that the quantum ergodicity theorem is 
equivalent to the semiclassical eigenfunction
hypothesis for ergodic systems for a subsequence of density one.

\end{appendix}

%%%%%%%%%%%%%%%%%%%%%%%%%%%%%%%%%%%%%%%%%%%%%%%%%%%%%%%%%%%%%%%%%%%%%%%%%%%%%

\renewcommand{\baselinestretch}{0.975} \small\normalsize

%\bibliographystyle{my_unsrt}
%\bibliographystyle{unsrt}

%\bibliography{biblio}

\end{document}